\documentclass[iop, numberedappendix, appendixfloats, tighten]{emulateapj}
\usepackage[]{times}
\usepackage[]{natbib}
\newcommand{\soho}{{\em SOHO{}}}
\newcommand{\trace}{{\em TRACE{}}}

\newcommand{\pref}{\protect\ref}

\begin{document}

\slugcomment{Accepted to Appear - Astrophysical Journal [Online material {\url http://tinyurl.com/29s7c4o}]}
\shorttitle{Upflows or Waves?}
\shortauthors{De~Pontieu \& McIntosh}
\title{Quasi-periodic Propagating Signals in the Solar Corona:\\The Signature of Magnetoacoustic Waves or High-Velocity Upflows?}

\author{Bart De Pontieu\altaffilmark{1,3}, Scott W. McIntosh\altaffilmark{2,3}} 
\altaffiltext{1}{Lockheed Martin Solar and Astrophysics Laboratory, 3251 Hanover St., Org. ADBS, Bldg. 252, Palo Alto, CA  94304}
\altaffiltext{2}{High Altitude Observatory, National Center for Atmospheric Research, P.O. Box 3000, Boulder, CO 80307}
\altaffiltext{3}{The authors contributed equally to the production of this manuscript.}
\email{bdp@lmsal.com,mscott@ucar.edu}

\begin{abstract}
Since the discovery of quasi-periodic propagating oscillations with periods of order three to ten minutes in coronal loops with {\em TRACE} and {\em SOHO}/EIT (and later with {\em STEREO}/EUVI and {\em Hinode}/EIS), they have been almost universally interpreted as evidence for propagating slow-mode magnetoacoustic waves in the low plasma $\beta$ coronal environment. Here we show that this interpretation is not unique, and that for coronal loops associated with plage regions (as opposed to sunspots), the presence of magneto-acoustic waves may not be the only cause for the observed quasi-periodicities. We focus instead on the ubiquitous, faint upflows at 50-150 km/s that were recently discovered as blueward asymmetries of spectral line profiles in footpoint regions of coronal loops, and as faint disturbances propagating along coronal loops in EUV/X-ray imaging timeseries. These faint upflows are most likely driven from below, and have been associated with chromospheric jets that are (partially) rapidly heated to coronal temperatures at low heights. 
These two scenarios (waves vs. flows) are difficult to differentiate using only imaging data, but careful analysis of spectral line profiles indicates that faint upflows are likely responsible for some of the observed quasi-periodic oscillatory signals in the corona. We show that recent EIS measurements of intensity and velocity oscillations of coronal lines (which had previously been interpreted as direct evidence for propagating waves) are actually accompanied by significant oscillations in the line width that are driven by a quasi-periodically varying component of emission in the blue wing of the line. This faint additional component of blue-shifted emission quasi-periodically modulates the peak intensity and line-centroid of a single Gaussian fit to the spectral profile with the same small amplitudes (respectively a few percent of background intensity, and a few km/s) that were previously used to infer the presence of slow mode magneto-acoustic waves.
Our results indicate that it is possible that a significant fraction of the quasi-periodicities observed with coronal imagers and spectrographs that have previously been interpreted as propagating magnetoacoustic waves, are instead caused by these upflows. The different physical cause for coronal oscillations would significantly impact the prospects of successful coronal seismology using propagating disturbances in coronal loops.
\end{abstract}


\section{Introduction}
Using observations from instruments on the {\em Solar and Heliospheric Observatory} \citep[\soho;][]{1995somi.book.....F}, the {\em Transition Region and Coronal Explorer} \citep[\trace;][]{1999SoPh..187..229H}, twin {\em STEREO} SECCHI/EUVI imagers \citep[][]{2008SSRv..136...67H}, and the Extreme-ultraviolet Imaging Spectrograph \citep[EIS;][]{2007SoPh..243...19C} on {\em Hinode} \citep[][]{2007SoPh..243....3K} the community has invested a great deal of effort in the identification and analysis of low-amplitude wave-like phenomena seen in EUV coronal imaging and spectroscopy \citep[see, e.g.,][for a few of the most cited, and recent examples]{2000A&A...355L..23D, 2002SoPh..209...61D, 2002SoPh..209...89D, 2003A&A...404L...1K,2005ApJ...624L..57A, 2008SoPh..252..321M, 2008A&A...482L...9O, 2008A&A...489.1307W, 2008ApJ...681L..41M, 2008A&A...491L...9V, 2008A&A...487L..17V, 2009ApJ...697.1674M, 2009ApJ...706L..76M, 2009A&A...503L..25W, 2009ApJ...696.1448W, 2010arXiv1003.0420M}. \citet{2005LRSP....2....3N} provides an excellent overview of the techniques used and highlights the community's interest in isolating and characterizing coronal waves, in order to remotely sense the physical attributes of the outer solar atmosphere by studying propagation speeds, amplitudes, and phase relationships of the observed phenomena.

\begin{figure*}
\epsscale{1}
\plotone{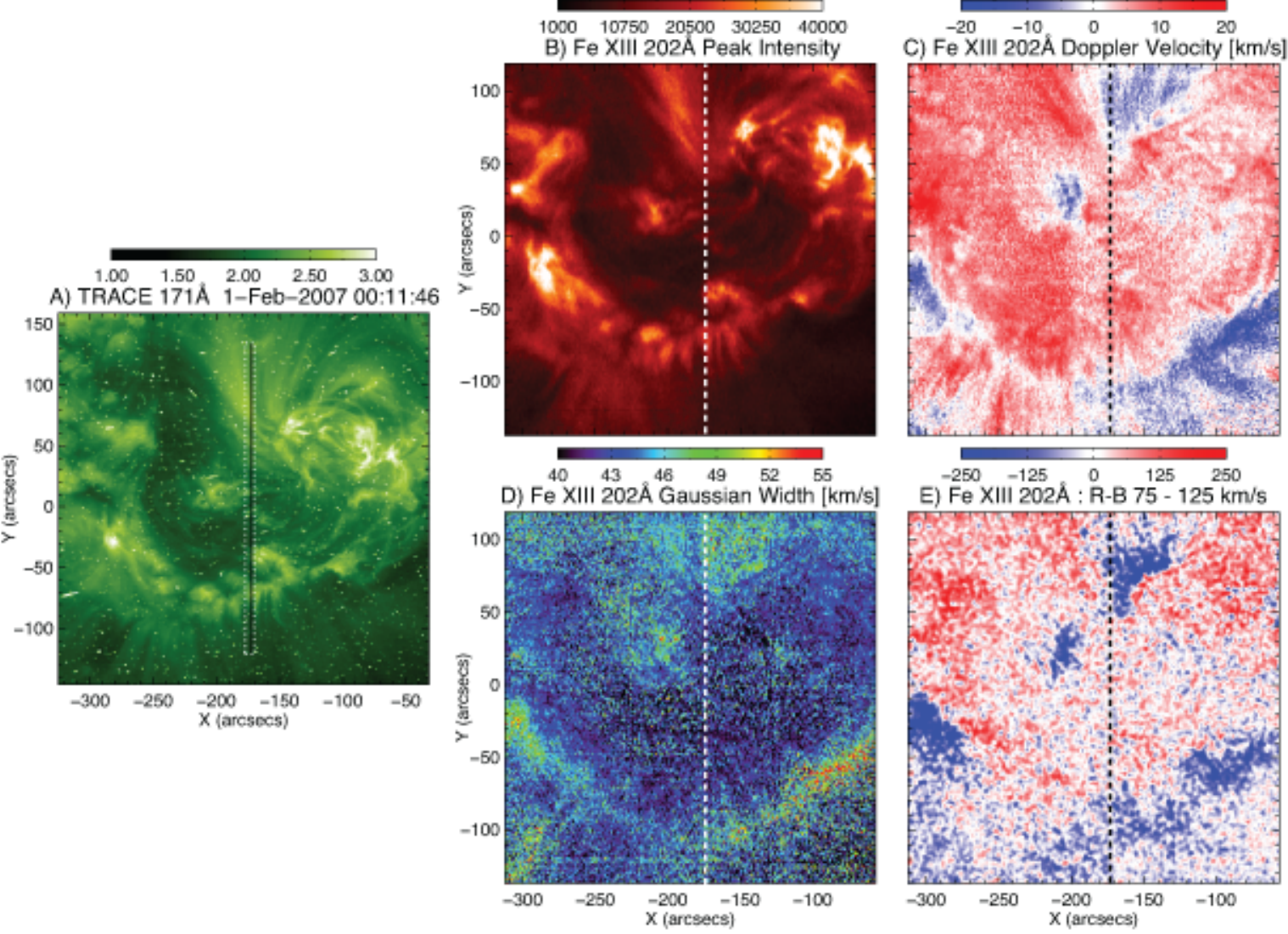}
\caption{{\em TRACE} 195\AA{} intensity (panel A), inferred single Gaussian fit parameters to the EIS spectroheliogram  showing the peak intensity (panel B), (relative) Doppler velocity (panel C), Gaussian width (panel D) and the results of the 75-125km/s R-B analysis in \ion{Fe}{13} 202\AA{}. On each of the panels we also show the pointing of the timeseries (vertical solid line), and a dashed box showing the maxima of the EIS pointing drift for the timeseries observation. Panels A and E of this figure are supported by movies in the online edition of the journal showing the pointing variation of EIS on the \trace{} images and the complete range of the R-B analysis. \label{f1}}
\end{figure*}

The broad community effort to probe the properties of the coronal plasma and magnetic field by attributing MHD wave properties to (quasi-) periodic propagating disturbances will be accelerated by the availability of considerably higher signal-to-noise, high cadence, high spatial resolution, multi-wavelength imaging provided by the Atmospheric Imaging Array on the {\em Solar Dynamics Observatory} ({\em SDO}). It is the availability of this new, complex, and rich data in concert with recent investigations of chromospheric-coronal coupling \citep[][]{2009ApJ...701L...1D,2009ApJ...706L..80M} that motivate the work presented and the plainly stated {\it caveat emptor} warning that it carries: ``not everything periodic in the outer atmosphere is evidence of a wave''. 

In some sense we are revisiting an issue that was triggered by the first observations of propagating disturbances with {\em TRACE} and \soho/EIT, with early reports suggesting that flows may explain some of the observed disturbances \citep[e.g.][]{1999SoPh..187..261S}. However, the presence of coherent 3 minute oscillations emanating from sunspots (clearly associated with umbral oscillations), and the lack of spectroscopic evidence for the required high-velocity ($\sim$50-100 km/s) line-of-sight flows, quickly led to the dominant interpretation of propagating slow mode magneto-acoustic waves \citep[e.g.,][]{2000A&A...355L..23D}. This interpretation ignores the fact that many of the observed propagating disturbances (especially in coronal loops emanating from quiet Sun network and active region plage) do not show evidence of significant quasi-periodic signals. More importantly, inspired by recent observations of high speed, but faint upflows in spectra of the TR and coronal footpoints of active region loops \citep[e.g.,][]{2008ApJ...678L..67H}, and of pervasive propagating disturbances in imaging data of coronal loops \citep[][]{2007Sci...318.1585S}, we discovered that faint upflows are ubiquitous in footpoints of coronal loops in active regions \citep{2009ApJ...701L...1D} and quiet Sun alike \citep{2009ApJ...707..524M}. Using {\em Hinode}/SOT-EIS-XRT data, we have suggested a direct link between faint blueward asymmetries in TR/coronal spectra at loop footpoints, propagating disturbances in images of coronal loops, and often quasi-periodically recurring chromospheric spicules \citep{2009ApJ...701L...1D,2009ApJ...706L..80M}. Armed with these new findings, we use EIS spectra (described in \S~2) and re-analyze them with two novel analysis techniques: asymmetry analysis (\S~3), and guided double fits using a genetic algorithm (\S~4). We use Monte Carlo simulations to show (\S~5) how the observed variations of line intensity, centroid and line width\footnote{Hereafter we will simple refer to the line width as the term for the Gaussian width ($\sigma$) of the line profile - $I(\lambda,t) = I_0 e^{-(\lambda-\lambda_0)^2 / (2\sigma_0^2)}$.} are fully compatible with the presence of a faint, strongly blueshifted component that quasi-periodically recurs. We discuss the interpretation and impact of our findings in \S~6. We show that the complexities of the atmospheric coupling, while appearing to detract from ``coronal-seismology", offers the potential for a deeper understanding of the mass and energy flow between chromosphere and corona. 

\section{Observations \& Analysis}

The primary data set investigated is a combination of \trace{} and {\em Hinode}/EIS observations of active region AR10940 on February 1, 2007. Some of this data was discussed by \citet{2009A&A...503L..25W}. While the \trace{} observations, taken in the 195\AA{} passband (at a cadence of 74s and exposure times of 65s) are impacted by cosmic ray hits they provide very useful context for the interpretation of the EIS observations. For both instruments we perform the usual post-processing: correcting for cosmic ray hits, hot pixels, detector bias, and dark current, and, for EIS, converting the data numbers to physical intensities (in erg~cm$^{-2}$~s$^{-1}$~sr$^{-1}$~\AA$^{-1}$). In addition to the regular preparation we use the solarsoft routine ${\tt eis\_jitter.pro}$ to compute the pointing jitter of EIS using the spacecraft housekeeping data. As a result we can reduce the impact of the pointing jitter of the spectroheliogram and timeseries observations, and co-align \trace{} and EIS.

\begin{figure}[h]
\plotone{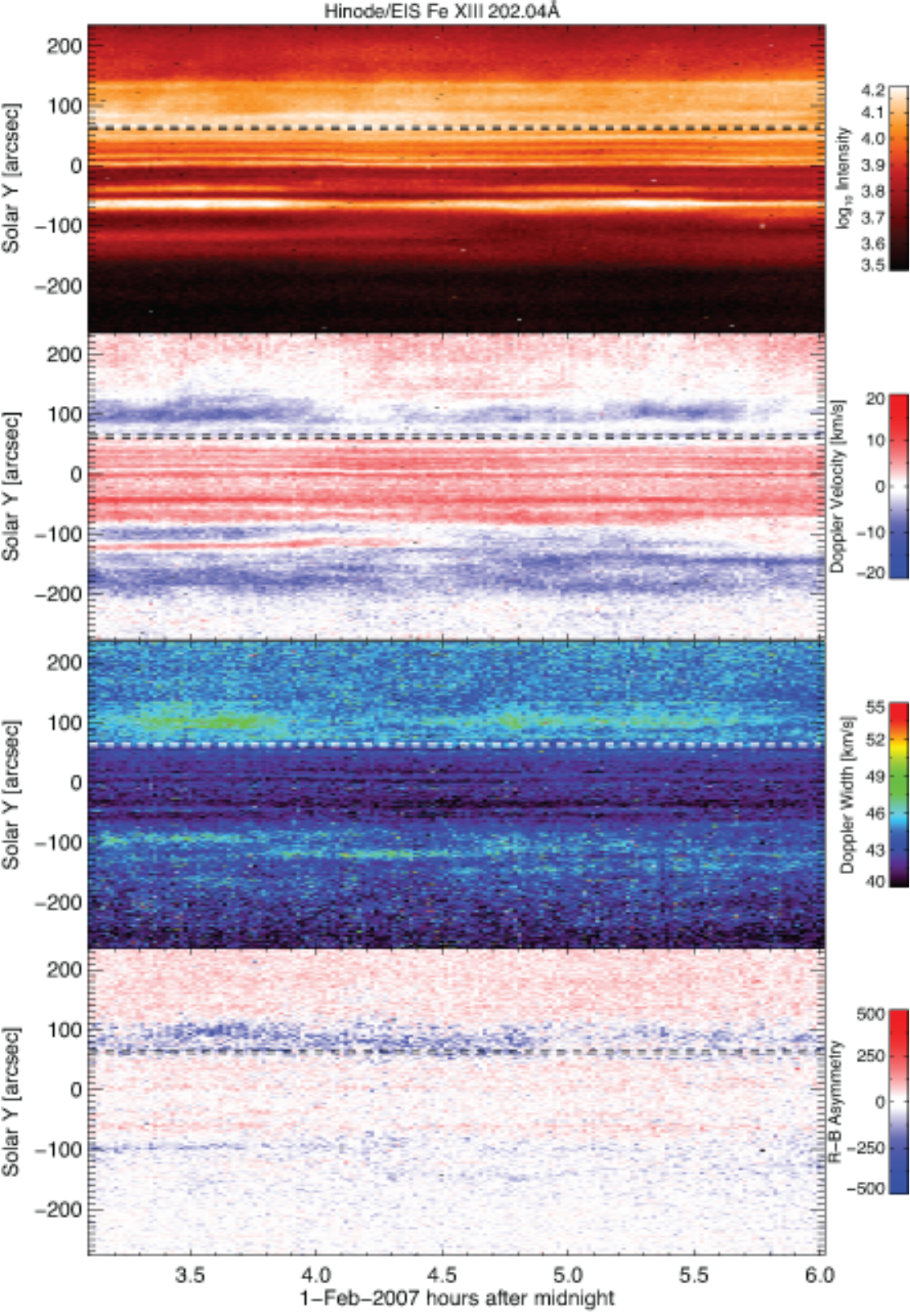}
\caption{Single Gaussian fit parameters to the EIS timeseries. From top to bottom: the peak intensity, (relative) Doppler velocity, width, and the results of the 75-125km/s R-B analysis in \ion{Fe}{13} 202\AA{}. On each of the panels we also show the region (y = 60-65\arcsec) along the slit chosen for further analysis.\label{f2}}
\end{figure}

There are two different types of EIS observations: the first is an 256\arcsec{} $\times$ 256\arcsec{} (West to East) spectroheliogram of the region starting at 00:12UT with a 60s exposure time at each rasterstep. The second part is a ``sit-and-stare'' (within the limits of the pointing jitter) observation starting at 01:32UT that involves 350 steps of the 1\arcsec{} $\times$ 512\arcsec{} slit with 60s exposures and solar rotation tracking at a median pointing of -140\arcsec{} ($\pm$24), -5\arcsec{} ($\pm$256). Panels A through D of Fig.~\pref{f1} provide the imaging context for the timeseries observation: Panel A shows the \trace{} 195\AA{} image at the start of the spectroheliogram observation with the EIS timeseries slit position (vertical thin line) and maxima of the pointing jitter drift shown as a dotted box; Panels B, C, and D show the peak line intensity, (relative) Doppler velocity and width that result from a single Gaussian fit to the spectra of the \ion{Fe}{13} 202\AA{} line which is formed (under equilibrium conditions) at $\sim$1.2MK \citep[][]{1998A&AS..133..403M}. In this case we choose to analyze the \ion{Fe}{13} 202\AA{} line instead of the \ion{Fe}{12} 195\AA{} line used by \citet{2009A&A...503L..25W} because the latter is affected by spectroscopic blends \citep[e.g.,][]{2008ApJS..176..511B}. This means that single Gaussian spectral fitting and studies of profile asymmetry are more reliable for the \ion{Fe}{13} 202\AA{} line \citep[e.g.,][]{2009ApJ...706L..80M, 2010arXiv1001.2022M} than for the \ion{Fe}{12} 195\AA{} line. The similar formation temperatures of both lines mean that we can reliably compare our results with those of \citet{2009A&A...503L..25W}. We will also exploit the spectroheliogram and time series observations in the \ion{Fe}{14} 264 and 274\AA{} lines \citep[formed at $\sim$2.5MK, ][]{1998A&AS..133..403M} to demonstrate, using more spectrally clean lines with relatively high signal-to-noise (S/N), that the observed phenomena are not isolated to a narrow temperature range. Unfortunately, while we observe similar diagnostics and variability in other clean, cooler spectral lines in these data (e.g., \ion{Si}{7} 275\AA) they have lower S/N and do not make for the most appropriate demonstration and so their presentation is omitted here.

\begin{figure}
\epsscale{1.15}
\plotone{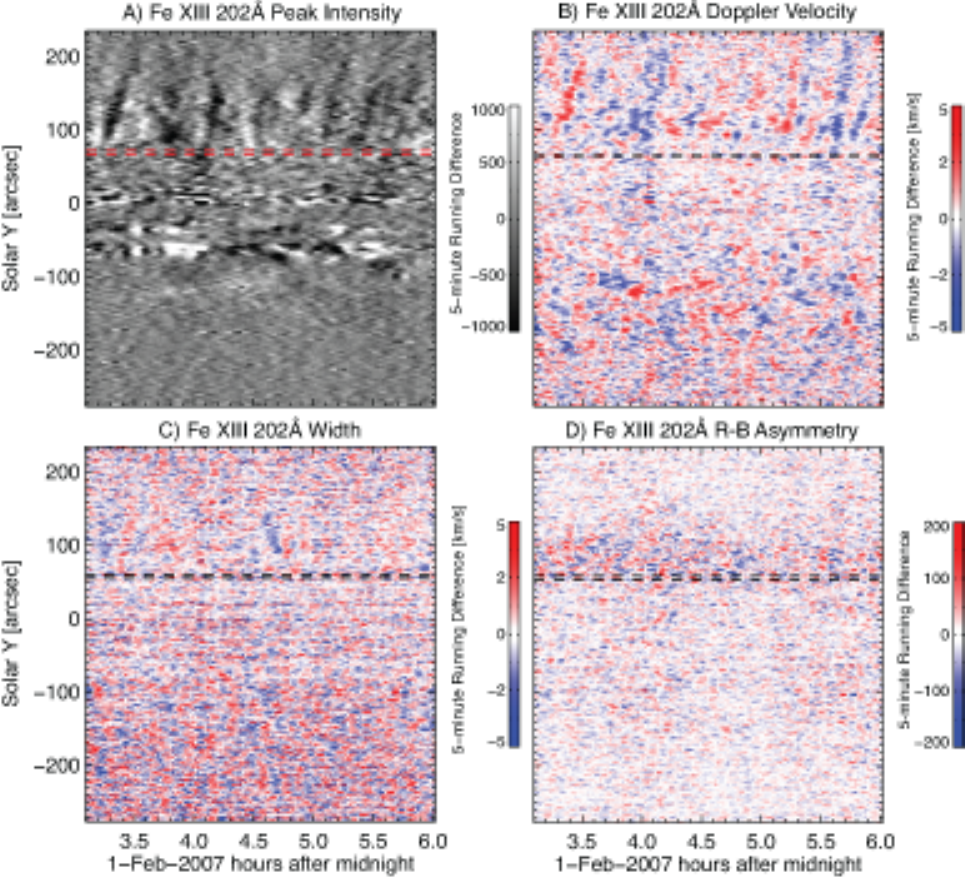}
\caption{Five-minute running differences for the EIS timeseries shown in Fig.~\pref{f2}. From top to bottom: the peak intensity, (relative) Doppler velocity, width, and the results of the 75-125km/s R-B analysis in \ion{Fe}{13} 202\AA{}.\label{f3}}
\end{figure}

\subsection{Line Profile Asymmetry Analysis}
Following the description of \citet{2009ApJ...701L...1D} and \citet{2009ApJ...706L..80M} we perform a `Red Minus Blue' (R-B) profile asymmetry analysis on spectral lines in the EIS data that are not significantly impacted by spectral blends in the relatively narrow (24 pixel) spectral windows used, e.g., the \ion{Fe}{13} 202\AA{} line. The R-B analysis involves several steps. First we fit a single Gaussian to the emission line profile to establish the line centroid. Once determined, we sum the amount of emission in narrow ($\sim$24~km~s$^{-1}$ wide) spectral regions symmetrically placed about the determined centroid. When summing we use a line profile that is interpolated to a spectral resolution that is ten times finer than that of EIS. We then subtract the red and blue wing contributions of the interpolated profile to make a filtergram that samples a particular velocity range. A positive value of R-B indicates an asymmetry in the red wing of the line, which we can interpret as the signature of excess down-flowing material at a velocity of that order while, conversely, a negative value of R-B would indicate an excess of up-flowing material. Panel E of Fig~1 shows the R-B diagnostic for \ion{Fe}{13} 202\AA{} at 75-125km/s and we are immediately drawn to the several locations where significant high velocity upflow signatures exist. Some of these underly the slit position of the time-series. Visual inspection of the TRACE movie accompanying panel A of Fig.~\pref{f1} shows that this is a location of propagating intensity disturbances (or so-called ``blobs''), exactly like those discussed in \citet{2009ApJ...706L..80M} \citep[or][]{2007Sci...318.1585S}. Indeed, the analysis of \citet{2009A&A...503L..25W} shows that the amplitude of the apparent motion of these blobs (deduced from motion along the loop in the plane of the sky) is of the same order as the velocity range that we have chosen for the R-B map shown.

\begin{figure}
\epsscale{1.15}
\plotone{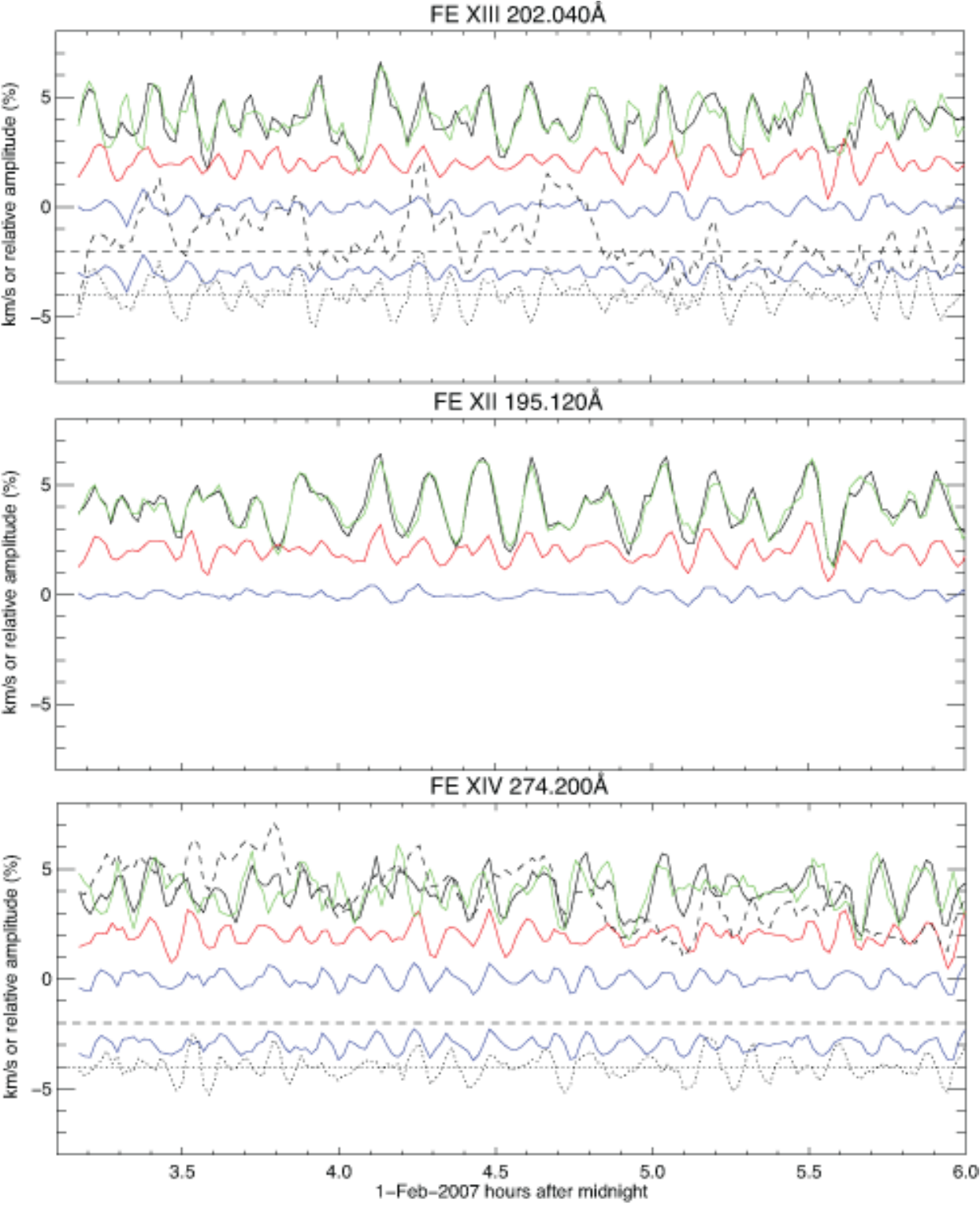}
\caption{Extracted EIS timeseries for the \ion{Fe}{13} 202\AA{} (top), \ion{Fe}{12}195\AA{} (middle) and \ion{Fe}{14} 274\AA{} (bottom) emission lines from 03:06-06:00UT, for locations $y=60-65\arcsec$. In each panel we show the detrended percentage change in signal amplitude or velocity change in km/s. The green trace shows the temporal variance in the peak line intensity while the black, red, and blue traces show the changes in total profile intensity, Doppler velocity (line center shift), and line width respectively. Compare and contrast with the black dashed and dotted lines which show the 74-157km/s R-B analysis and detrended form (as for V and V$_{nt}$). The zero lines for these are drawn at -2 and -4 respectively and to stress our convention that {\em values above that zero line represent an excess in the blue wing of the line}. For clarity, the timeseries for intensity is offset by +4, velocity by +2 and the bottom linewidth by -2 from the zero-point on the y-axis. \label{f4}}
\end{figure}

\section{Time-Series Analysis}
To facilitate comparison, we select the same portion of the de-jittered timeseries as that chosen by \citet{2009A&A...503L..25W}, i.e., from 03:06-06:00UT. Figure~\pref{f2} shows the single Gaussian fit parameters for the \ion{Fe}{13} 202\AA{} timeseries observations, from top to bottom showing the peak line intensity, (relative) Doppler velocity inferred from the shift of the profile centroid, the width and the 75-125km/s R-B analysis. We note the presence of a large region of blue-wing asymmetry in this line at positions $\sim$ y=-90\arcsec{} and around 60$\le$y$\le$110\arcsec{} that are consistent with locations identified in panel E of Fig.~\pref{f1}. The latter location is at the footpoint region of a loop fan that is clearly visible in the TRACE~195\AA{} image in Fig.~\pref{f1}, and on which blobs are seen to propagate in the {\em TRACE} and EIS data. The pointing jitter in the direction perpendicular to the slit most likely leads to the gradual, but significant decrease of the RB signal towards the end of the timeseries. We will discuss the effects of this pointing jitter further in \S~6.
\begin{figure*}
\epsscale{1}
\plotone{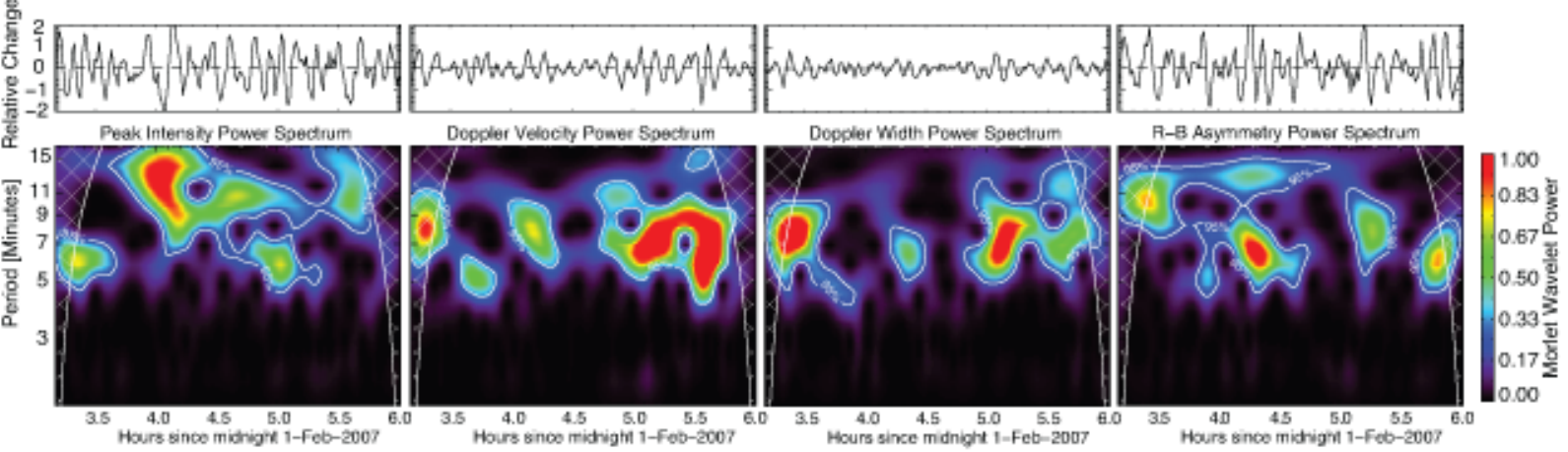}
\caption{Wavelet power spectra for the \ion{Fe}{13} 202\AA{} timeseries shown in the top panel of Fig.~\pref{f3}. The solid white contours denote regions of the wavelet power spectrum of 95\% significance and the cross-hatched region encloses the cone-of-influence for the power spectrum. \label{f5}}
\end{figure*}

A closer inspection of the peak intensity, the first and second moments of the Gaussian fit (i.e, Doppler velocity and line width) reveal episodes of small amplitude oscillations of a few percent in intensity and a few km/s amplitude for both the Doppler velocity and width. The oscillations of the peak intensity and Doppler velocity were reported previously by \citet{2009A&A...503L..25W} for the blended \ion{Fe}{12} 195\AA{} (cf. their Fig.~2 where, using running differences and detrending, they are highlighted beautifully). Here we highlight the (previously unreported) existence of oscillations in the line width that are in phase with those of the peak intensity and velocity, and that are accompanied by oscillatory changes in the R-B signal, with blueward asymmetries occurring in phase with line width increases. The oscillations in the line widths and R-B signal are weak, but visible in the 5-minute running differences shown in Fig.~\pref{f3} although with substantially lower contrast than the lower order Doppler velocity and line intensity measurements\footnote{The reduced contrast for higher order moments is understandable given the impact of photon noise on these measurements. This is illustrated in detail in \S~4 with Monte Carlo simulations.}. The sign-flipping diagonal striations in the space-time plots, especially those around 60\arcsec{} $\le$ y $\le$ 150\arcsec{} of all four measurables are clearly related. As we will illustrate below, these line width and blueward asymmetry oscillations are very valuable diagnostics that provide strong physical constraints on what is driving the low amplitude, quasi-periodic, signal in the data. 

The correlation between intensity, velocity, line width and blueward asymmetry is more clearly illustrated in Fig.~\pref{f4}. This figure shows an example oscillatory region from the timeseries at the y=60-65\arcsec{} location as shown by the horizontal lines in Fig.~\pref{f2}. We have summed over the same region that was studied by \citet{2009A&A...503L..25W} in their Fig. A2. In the left panel of Fig.~\pref{f4} we use the recipe\footnote{$I_{detrended}= (I_{peak} - <I_{peak}>)/<I_{peak}>$ with $<I_{peak}>$ the running average over 10 minutes. For velocities and width, we show $v_{detrended} = v - <v>$, with $<v>$ the running average over 10 minutes. } of \citet{2009A&A...503L..25W} and show the resulting, detrended variations in the peak intensity (black solid line), total intensity (here defined as the peak intensity times the line width; green solid line), Doppler velocity (red solid line), Gaussian line width (blue solid line), the R-B measure for velocities of 75-125km/s (black dashed line) and its detrended form (black dotted line). We repeat this process for the \ion{Fe}{12} 195 \AA{} line (middle) and the \ion{Fe}{14} 274\AA{} line (bottom), to illustrate that these variations are not isolated to the formation temperature of \ion{Fe}{13}. Again, although not shown here, analysis of \ion{Si}{10} 261\AA{} and \ion{Si}{7} 275\AA{} reveal a similar behavior, but are omitted here because those lines have significantly lower S/N. We also note that the presence of a slight blend (and/or slight gradient in the background emission) in the red wing of \ion{Fe}{13} leads to occasional more slowly evolving excursions of the absolute RB measure towards the red. Detailed analysis suggests that the blend is likely caused by a much hotter line. The decent correlation between the RB asymmetry and the other moments of the \ion{Fe}{13}, and with the RB measure in other lines (\ion{Fe}{14} and \ion{Si}{10}) indicates that the detrended RB measure is still a good measure of blueward excursions, especially since the redward excursions happen on longer timescales than the oscillations we focus on here. 
 
Detailed analysis of the \ion{Fe}{13} and \ion{Fe}{14} panels in Fig.~\pref{f4} show that the correlation between intensity, velocity, line width and blueward asymmetry is by no means perfect, but significant throughout the 3 hour long timeseries. Most of the line width peaks are associated with stronger blueward asymmetries and with blueward excursions of the line centroid, and intensity peaks. These correlations will be explored further in a statistical sense in \S~4. We also note that the results of Fig.~\pref{f4} are by no means unique: we have found many other locations in the same dataset that show such correlations. Two examples are shown in the appendix (Figs.~\pref{f4b}, \pref{f4c}). 

Detailed comparison of the time variations of these lines also shows that the oscillations of the intensity and Doppler velocity of \ion{Fe}{13} 202\AA{} and \ion{Fe}{12} 195 \AA{} have similar amplitudes and are very well correlated. While the correlation with \ion{Fe}{14} 274\AA{} is not as clean, it is significant. This suggests that the driving mechanism for these oscillations acts over a wide range of temperatures. This does not seem compatible with a scenario in which sound waves propagate through an isothermal medium, as suggested for other datasets \citep{2009ApJ...706L..76M}. 

\begin{figure}
\epsscale{1.15}
\plotone{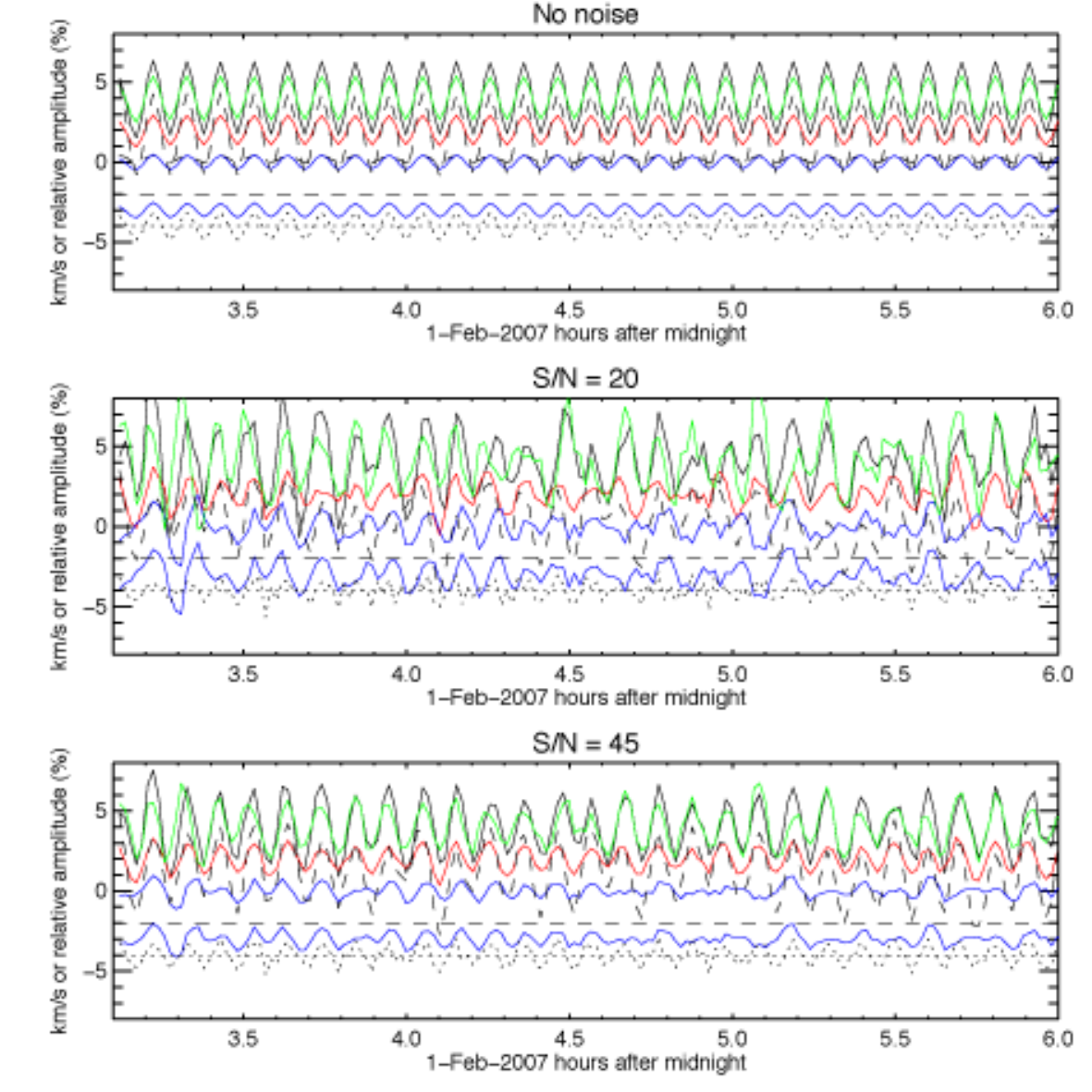}
\caption{Simulated timeseries of spectral line parameters from a forward model that simulates the effect of a faint and periodic secondary Gaussian component on the total intensity (full black line), peak intensity (green), line centroid (red) and line width (blue) from a single Gaussian fit to the total spectral line profile (which is the sum of two Gaussians reduced to EIS spectral resolution at 202\AA). The relative amplitude of the secondary component is 5\% of the stationary background component, with $\sigma_1=\sigma_2=30$ km/s, and the second component is offset $\delta v = 50$ km/s to the blue of the dominant component. We use the same convention as in Fig.~\pref{f4}: in each panel we show the detrended percentage change in signal amplitude or velocity change in km/s. The black dashed and dotted lines show the 55-155km/s R-B analysis and detrended form (same recipe as for line centroid and width). The zero lines for these are drawn at -2 and -4 respectively and to stress our convention that {\em values above that zero line represent an excess in the blue wing of the line}. For clarity, the timeseries for intensity is offset by +4, velocity by +2 and the bottom linewidth by -2 from the zero-point on the y-axis.
The input periodic signal has a period of $P/2=6$ minutes. The upper panel assumes that there is no photon noise, the middle and bottom panels have the same input periodic secondary component, but with photon noise added. The middle panel is for $S/N=20$, which is similar to the quality of a measurement of an individual EIS pixel in the dataset studied (1-Feb-2007), whereas the bottom panel is for $S/N=45$, which is representative of the average over 5 pixels that is shown in Fig.~\pref{f4}. The simulated profiles (with noise added) agree very well with the data of Fig.~\pref{f4}, both with respect to the average amplitude of the variations, and the less-than-ideal correlations between some of the parameters. The agreement is remarkable when taking into account the fact that the Sun most likely does not produce a secondary component that is perfectly periodic and has a constant amplitude.\label{f11}}
\end{figure}

The temporal behavior of these single Gaussian fit parameters strongly suggests quasi-periodic behavior. The wavelet analysis \citep[e.g.,][]{1998BAMS...79...61T} presented in Fig.~\pref{f5} show, from left to right, the normalized wavelet power spectra of the intensity, doppler velocity, line width, and R-B asymmetry timeseries of Fig.~\pref{f4}. Outside the Cone-of-Influence (the white cross-hatched regions) and inside the 95\% signal significance levels (solid white contours that are estimated using a red noise background) we see that periodicities occur throughout the 3 hour long timeseries. These timeseries are best described as quasi-periodic with periods in a range from five to twelve minutes dominating - similar to those reported for \ion{Fe}{12} 195\AA\  \citep{2009A&A...503L..25W}. This confirms the similarity in the oscillatory behavior of the \ion{Fe}{13} and \ion{Fe}{12} lines. The quasi-periodicities in the R-B asymmetry suggests that the blueward asymmetries and associated upflows recur quasi-periodically.

\section{Forward Modeling}

What causes these oscillations? The close correlation between blueward asymmetries and increased line width (as well as blueshifted line centroid and increased peak intensity) shown in Fig.~\pref{f4} provides a strong clue. We use forward modeling to show that these observations are fully compatible with a scenario in which the quasi-periodic occurrence of a strongly blueshifted, but faint emission component causes a blueward asymmetry (R-B signal) at high velocity, and various changes in the line parameters deduced from a {\em single} Gaussian fit to the spectral profile (as is common practice): slight increases in the line width, the position of the line center (the Doppler velocity), and the peak intensity of the line.

\begin{figure*}
\epsscale{1}
\plotone{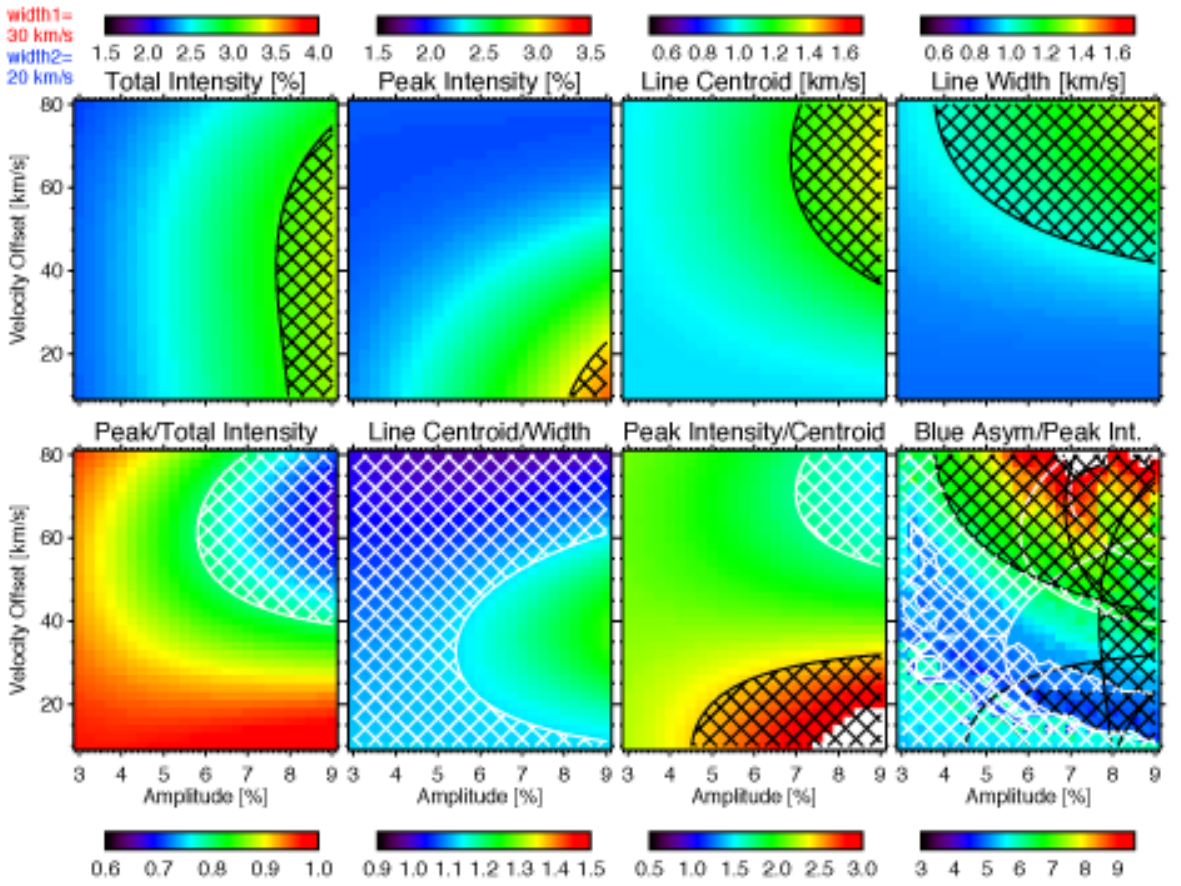}
\caption{Results of a parameter survey of the forward model in which we investigate the impact (on single Gaussian fit parameters) from a periodically recurring secondary component superimposed on a stationary background component. The properties of the two Gaussian components to the spectral line profile are varied over a wide range: $0.03<a<0.09$ and $\delta v = 10-80$ km/s, with fixed $\sigma_0=30$ km/s and $\sigma_1=20$ km/s. The secondary component is periodically recurring with a period $P/2=6$ minutes. The top row shows, as a function of $a$ (amplitude) and $\delta v$ (velocity offset), the standard deviation over a one hour time range for the following fit parameters from a single Gaussian fit to the spectral line profile that is reduced to EIS spectral resolution: total intensity (defined here as peak intensity times line width), peak intensity, line centroid and line width. The bottom row shows the ratio between peak and total intensity, line centroid to line width, peak intensity to line centroid, and blueward asymmetry to peak intensity. The contours show the range of parameters that are ``allowed'' based on measurements with EIS (see text for details), with black and white contours respectively for upper and lower values of the allowable range. The hashed out regions show the non-allowed values. The bottom right panel shows the combination of all observational constraints, with only a small region for amplitudes of $4-8$\% and offset velocities of order 30-50 km/s. An accompanying movie (available in the online edition of the journal) illustrates that while the shape of the region of allowed parameters changes somewhat for different values of $\sigma$, the range remains locked around amplitudes of order 3-8\% and velocities of 30-50 km/s.
 \label{f9}}
\end{figure*}

To investigate this scenario, we use Monte Carlo simulations in which we construct a timeseries of simulated spectra with properties similar to the EIS observations. First, we calculate for each timestep the emergent spectrum assuming the spectral line emission is dominated by two different sources, both with Gaussian profiles (as a function of wavelength). The spectral line profile is given by the sum of two Gaussians:

\begin{equation}
  I(\lambda,t) = I_0 e^{-(\lambda-\lambda_0)^2 / (2\sigma_0^2)} + \\
  a I_0 \, cos^2 (2\pi t/P) \, e^{-(\lambda-\lambda_1)^2/ (2\sigma_1^2)} 
\end{equation}

where $I_0$, $\lambda_0$, and $\sigma_0$ are, respectively, the intensity, constant central wavelength, and line width of the steady background component, and where $a$ ($<<1$) is the relative amplitude of the fainter second component compared to the background component, $\lambda_1$ and $\sigma_1$ are the center position and  line width of the second component. The intensity of the second component is allowed to change with time (with a period $P/2$). This represents the recurring upflowing component that modifies the signal with time.

To obtain EIS-like spectra, we convolve $I(\lambda,t)$ with a Gaussian that has a (Gaussian) line width equal to the instrumental broadening of EIS (22.93~m\AA). Next, we rebin the high resolution spectra to EIS resolution (with one EIS pixel equal to 22.93~m\AA). For each timestep we add photon noise ($\sqrt{I(\lambda,t)}$) to each spectral pixel using random numbers from a Gaussian distribution given by the IDL function {$\tt randomn.pro$}. Next, we use {$\tt gaussfit.pro$} to fit a single Gaussian to the emergent profile, and detrend the synthetic time series in a fashion identical to that used to prepare Fig.~\pref{f4}. Finally, we perform an R-B asymmetry analysis identical to that performed earlier on the line profiles observed with EIS (see \S~2 for details).

We repeat this recipe for a wide range of values for $a$, $\sigma_0$,  $\sigma_1$, and $\delta v [= c (\lambda_0-\lambda_1)/\lambda_0]$; where $c$ the speed of light]. An example timeseries for the peak intensity, total intensity, line centroid, line width and R-B asymmetry analysis for $\delta v = 50$ km/s, $a=0.05$ and $\sigma_0=\sigma_1=30$ km/s is shown in Fig.~\pref{f11}. The top row shows, for this case, and when no noise is present, that the oscillations in peak and total intensity are of order a few percent, whereas the oscillations in line centroid and width are of order 1 km/s. All of these oscillations are in phase with each other, with the driver (the periodically recurring faint second component), and with oscillations in the R-B asymmetry: peaks in intensity occur at the same time as blueward excursions of the line centroid, the peaks in line width and the blueward asymmetries. To allow direct comparison we have produced plots that are identical in range in x and y to Fig.~\pref{f4}. Comparing the top row of Fig.~\pref{f11} and Fig.~\pref{f4} shows that generally the appearance of the ``no noise'' case is similar to what we observe with EIS. 


This correspondence (in a statistical sense) is even more striking when we compare Fig.~\pref{f4} with the two other rows of Fig.~\pref{f11}. We have calculated this particular realization for three different levels of signal to noise (S/N): infinite (no noise), S/N=20, and S/N=45. Using the errors calculated by {$\tt eis\_prep.pro$}, we estimate that single pixels in the region of the observed oscillations (in the vicinity of $y=60-65\arcsec$) have a S/N of order 20. However, the oscillations shown in Fig.~\pref{f4} were based on summing over 6 spatial pixels and thus have a larger S/N of order 45. The summing in this region was performed to allow direct comparison with the results of \citet{2009A&A...503L..25W} who suggested that the oscillations may be coherent in this region. We see that the $S/N=45$ case shows an amazing correspondence (in a statistical sense) with the oscillations observed with EIS in the \ion{Fe}{13} 202\AA\ line. This agreement is even more remarkable given the fact that the Sun most likely does not produce a perfectly periodic, constant amplitude secondary component. We also note that the increased noise-level has the strongest impact (not surprisingly) on the higher order moments of the profile: line centroid and width. It also has a significant effect on the R-B measure, and a somewhat smaller effect on the peak intensity. The increased noise leads to a correlation that is significantly worse than the ``no-noise'' case: random phase shifts between parameters occur, peaks are occasionally replaced by troughs in one parameter and not the other, and the amplitude of the oscillations in line centroid and width (and thus ``total'' intensity) oscillations are significantly reduced - these effects are also clearly present in the EIS data shown in Fig.~\pref{f4}. 

The results of the Monte Carlo simulations thus strongly indicate that the observed oscillations with EIS are  compatible with a double component plasma where there is a quasi-periodically recurring upflowing component at 5\% of the stationary component's intensity, and with a relative velocity of order 50 km/s. This provides evidence for a scenario in which upflows cause some of the observed oscillations in intensity (e.g., in the {\em SOHO}, {\em TRACE}, {\em STEREO}, and EIS imaging data) and velocity and line width (in CDS and EIS spectra). Our results also indicate that the observed, less-than-ideal, correlations between the parameters are not a sign of a lack of correlation, but instead are exactly what is expected from the significant impact of finite spectral resolution, instrumental broadening and photon noise in the EIS instrument. We note that the line widths we have assumed (20-30 km/s) are similar to what is observed at the locations where EIS observes oscillations (after subtracting instrumental broadening). 

How unique is this interpretation? How well can we constrain the properties of the second, faint, component? Are there other combinations of two Gaussian profiles (with one periodically recurring) that would lead to the same oscillations in line intensity, centroid and width? To address this issue, we performed a parameter search in 4 dimensions for the case of profiles with S/N=30: varying $a$ from values of 3 to 9\%, $\delta v$ from 10 to 80 km/s, and $\sigma_0$ and $\sigma_1$ independently from 20 to 40 km/s. For each of the resultant timeseries of line (peak and total) intensity, centroid and width, we calculate the standard deviation over the course of one hour. We also calculate the ratio of the standard deviation of the peak intensity to that of the total intensity, and similar ratios for centroid to width and peak intensity to the centroid. We also calculate the standard deviation of the blueward asymmetry normalized to the peak intensity. For the sake of comparison we analyze the EIS observations of the locations at $y=60-65\arcsec$ and determine similar values for the observed oscillations. We use these observed standard deviations to constrain the parameter range of the forward models. This works surprisingly well, as shown in Fig.~\pref{f9}. By combining all 4 spectral line parameters with well-chosen ratios of those parameters, we find that for a given value of $\sigma_0$ and $\sigma_1$, only a small range of amplitudes $a$ and velocity offsets $\delta v$ are compatible with the oscillations observed with EIS. Fig.~\pref{f9} shows that for $\sigma_0=30$ km/s, and $\sigma_1=20$ km/s, only second components with amplitudes of order 4-8\% and velocity offsets of order 30-50 km/s (blueward of the background component) are compatible with the observed behavior. This region is shown as the only region in the bottom right panel that is not hashed out. All other regions are excluded by constraints imposed by the EIS observations. 

As one might expect, these results depend significantly on the signal to noise assumed. For example, lower S/N observations will naturally lead to relatively higher intensity perturbations as a direct consequence of photon noise. However, we have made sure that our assumed S/N of 30 matches that of the S/N of the observed constraints (derived from measurements based on summing over 2 spatial pixels, with each individual pixel having a S/N of $\sim$20).

The results also depend on the values assumed for $\sigma_0$ and $\sigma_1$. An accompanying movie (available in the online edition of the journal) illustrates that while the shape of the region of allowed parameters changes somewhat for different values of $\sigma$, the range remains locked around amplitudes of order 3-8\% and velocities of 30-50 km/s. In fact, we find that for widths less than 20 km/s and more than 35 km/s, there are no solutions that are compatible with the observed oscillations (under our assumptions). Moreover, for realistic values of the line widths (which are of order 20-30 km/s after subtracting instrumental broadening, see also \S~5), there are {\em no} solutions for offset velocities of order 10 km/s (which would be necessary for the wave interpretation of \citet{2009A&A...503L..25W}, see \S~6).

The forward model presented here assumes that the upflowing component is perfectly periodic, with identical amplitude throughout the timeseries. In reality, the upflowing component is likely only quasi-periodic, with an amplitude that can vary erratically. As a result, the standard deviation of each parameter might be different, which could impact the range of allowable parameters. However, the most stringent constraints are actually the ratios of the standard deviation of various parameters. These ratios are unlikely to change much if the upflowing component were variable in time and period, since all parameters will be impacted equitably by a lack of oscillatory power. Nevertheless, because of some of these limitations it is prudent not overinterpret the forward model. Consequently, we conclude that a second component that has a {\em range} (as opposed to exact values) of parameters (3-8\%), and offset velocities (30-50 km/s) can explain the observed oscillations.

In summary, we find that the results of the forward model strongly support the scenario posed in this Paper, that of a faint, but quasi-periodically recurring upflow component which impacts the intensity, centroid and line width of a single Gaussian fit to the spectral line profile. 



\begin{figure}[h]
\epsscale{1.15}
\plotone{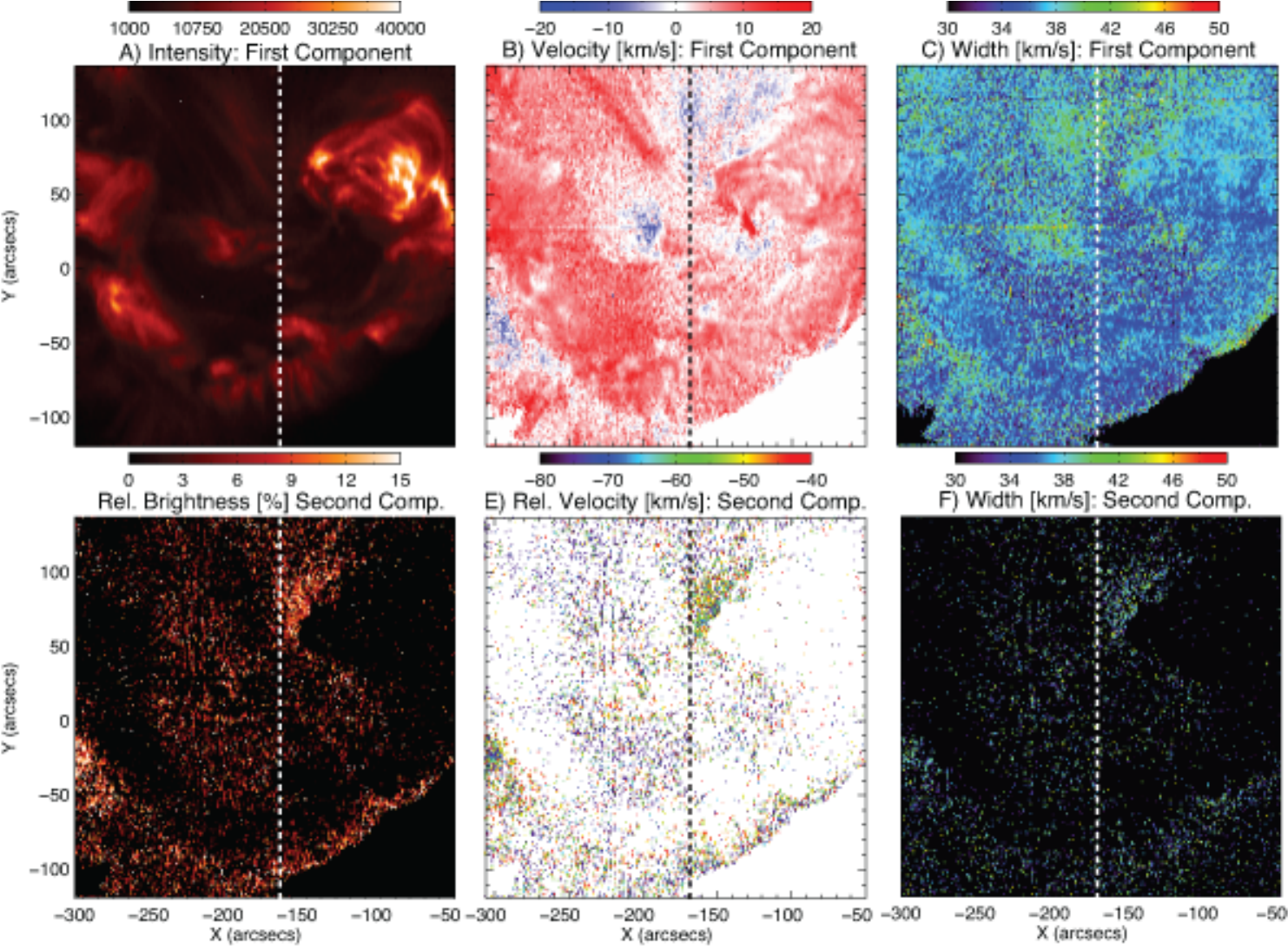}
\caption{Sample GA double Gaussian fits for the {\em Hinode}/EIS Spectroheliogram of Fig.~\pref{f1} in the \ion{Fe}{14} 274\AA{} emission line. From top to bottom, left to right, the figure shows the peak intensity, Doppler velocity, and line width of the core component, the relative brightness,  Doppler velocity and line width of the second component. \label{f6}}
\end{figure}

\section{R-B ``Guided'' Double Fits}
Can we find direct evidence of this second component by performing double Gaussian fits? Motivated by these Monte Carlo simulations, and the presence of the line profile asymmetries in multiple spectral lines we now look to see if we can {\em directly} characterize the additional emission component. To do this we explicitly choose a double Gaussian component fit as one that is minimally consistent with the data (of course more free parameters will ensure a ``better'' fit to the data). In this case we chose not to perform a ``blanket'' double Gaussian profile fit, that is one where the fit parameters are completely free, or one where a physical premise is used to deduce, or loosely impose, where the components lie in the spectral domain \citep[e.g.,][]{2000A&A...360..761P}. We take advantage of the fact that the line profile asymmetries and large line widths are co-spatial with locations where the quality of fit measure, (i.e., the reduced $\chi^{2}$) is anomalously large\footnote{If the $\chi^{2}$ map that results from the single Gaussian fit to the line profiles is not spatially ``flat'' then the spectra are not adequately described by a single Gaussian; i.e., there is something missing in the underlying physical description of the data \citep[e.g.,][]{2003drea.book.....B}.}. We use the presence of a R-B asymmetry in the line profile of a given pixel to determine that a double fit is needed, and to provide the fitting algorithm with starting parameters for that second component. For this first test of such a procedure we use a Genetic Algorithm (GA) Gaussian fitter that is based on a well tested, robust, although slow method \citep[][]{1998A&AS..132..145M}.

This guided double fit procedure is quite simple. We find all locations where the R-B asymmetry exceeds a given magnitude (in this case 1\% of the peak line intensity). We then approximate the R-B profile (as a function of velocity/wavelength) by a Gaussian, and use the Gaussian only to estimate the center (in wavelength) of the R-B enhancement. The centroid of the initial line profile fit (used to compute the original R-B asymmetry) and the centroid of the R-B asymmetry profile which, depending on the sign of the R-B asymmetry will be on the blue (negative) or red (positive) side of the initial (core) profile, are the {\em only} values supplied to the GA Gaussian double fit. This algorithm then undertakes a global minimization of the double Gaussian line spectra fit allowing both core and second component centroids to move by 1 spectral pixel to the blue or red of the supplied position. For reference, at 195\AA{} in EIS this would allow a shift of $\pm$35km/s from the input parameters for both emission components. We note also that all other parameters in the fit are free to span the range of possible values.

Figure~\pref{f6} shows the results of applying this technique to the spectro-heliogram shown in Fig.~\pref{f1} for the \ion{Fe}{14} 274\AA{} emission line. While rough, panels A, B, and C are consistent with their contemporaries in Fig.~\pref{f1} while panels D, E and, F show the relative brightness ($I_{Wing}/I_{Core}$) of the second (wing) emission component, its velocity relative to the first component, and the line width respectively. The regions of large profile asymmetry under the line EIS slit for the timeseries observation show a second emission component of approximate amplitude 6-8\%, a velocity of 45-65km/s and (Gaussian) line width of 25-35km/s\footnote{The values for the width in this section exclude the instrumental broadening, which has been subtracted to allow comparison with the input values for the forward model.}. The widths of the second and first component are similar.

\begin{figure}
\epsscale{1.15}
\plotone{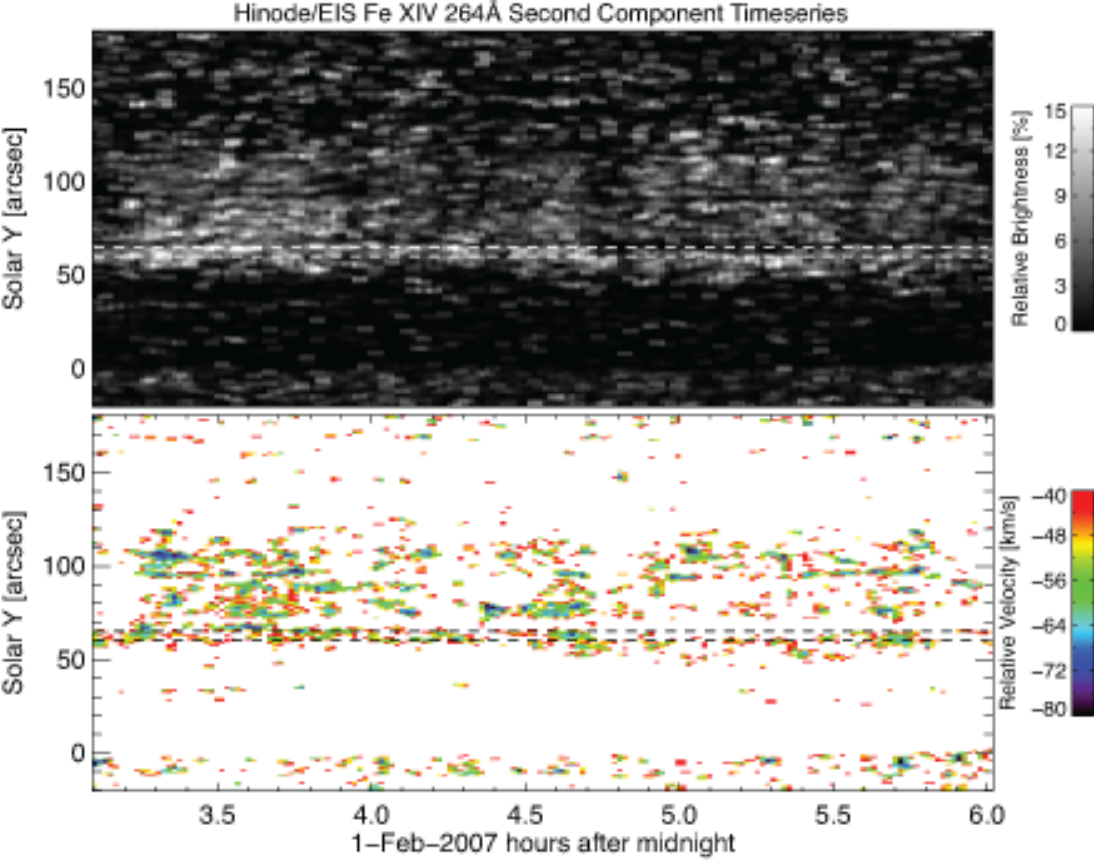}
\caption{Variation in the second GA double Gaussian fit components for the {\em Hinode}/EIS time series observation of Figs.~\pref{f2} and ~\pref{f3} in the \ion{Fe}{14} 264\AA{} emission line. The top panel shows the relative brightness ($I_{Wing}/I_{Core}$) of the second (wing) emission component while the bottom panel shows its Doppler velocity (relative to the core component). For reference the dashed horizontal lines show the location of the time series chosen for Fig.~\pref{f4}. \label{f7}}
\end{figure}

\begin{figure}
\epsscale{1.15}
\plotone{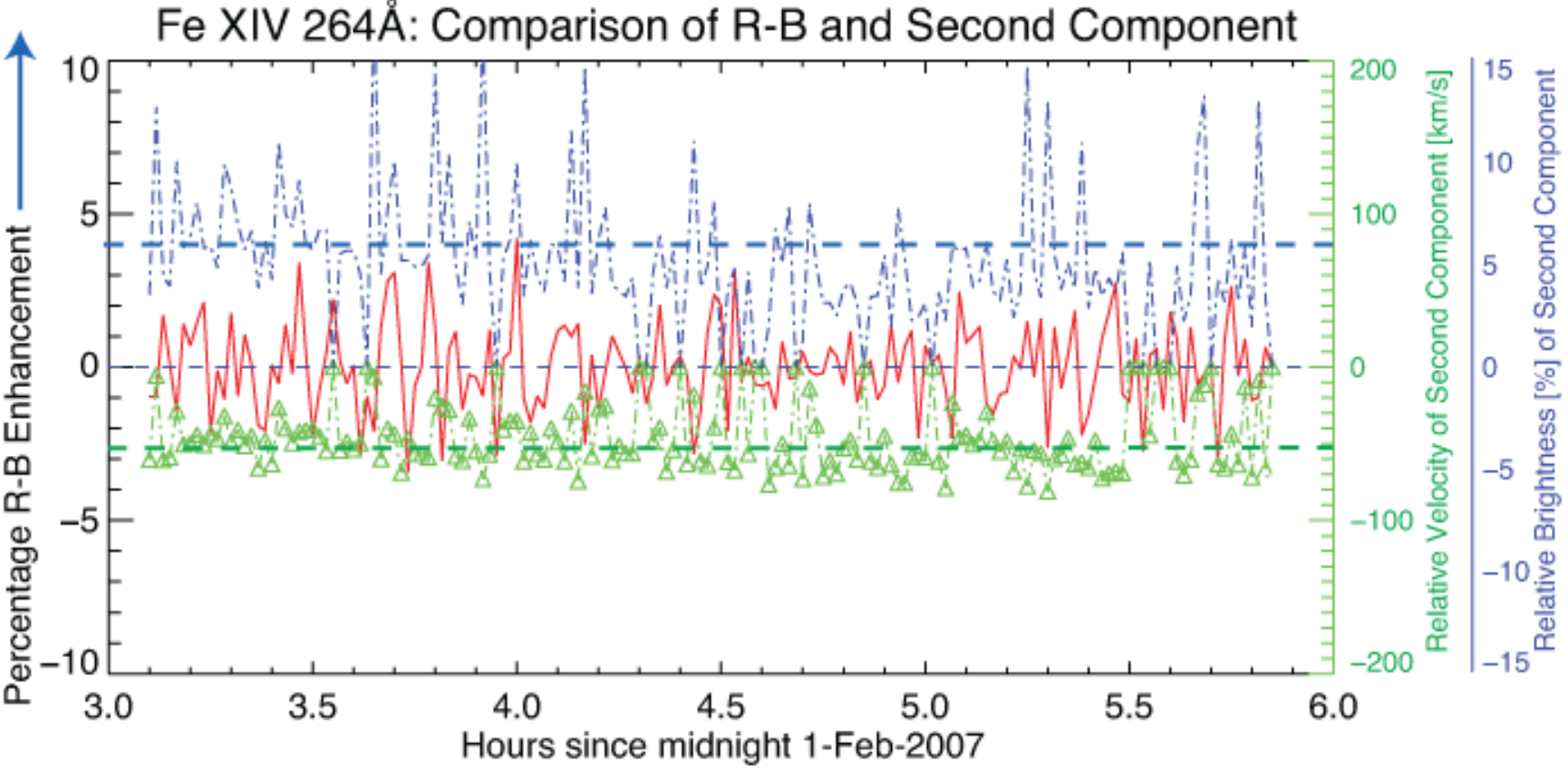}
\caption{Extracted variation of the second \ion{Fe}{14} 264\AA{} emission components of relative brightness (blue line) and velocity (green triangles) versus the variation in the (inverted) R-B diagnostic (red line, with blueward asymmetries positive). \label{f8}}
\end{figure}

Figure~\pref{f7} shows the results of applying the same R-B guided double Gaussian fit technique to the timeseries observations in the \ion{Fe}{14} 264\AA{} emission line. This line is formed at the same temperature (in equilibrium) as \ion{Fe}{14} 274\AA{} and is used here to illustrate that similar spectral asymmetries resulting from the second emission components occur across a range of temperatures. The two panels show the relative brightness ($I_{Wing}/I_{Core}$; top) of the second (wing) component and its position relative to the rest position of the line (bottom). We see that the intensity of the second component modulates, but that the velocity of the second component stays relatively stable at 35-60~km/s. The (Gaussian) line width of the second component (not shown) is also stable in magnitude at 25-35~km/s.

To illustrate these properties we show for \ion{Fe}{14} 264\AA{} (Fig.~\pref{f8}) the variation in the relative brightness (blue line) and velocity (green triangles) of the second component, extracted from the region inside the dashed lines (cf. Fig.~\pref{f2} and~\pref{f4}), with the percentage change in R-B assymmetry at the same location. As for Fig.~\pref{f4} the R-B timeseries is inverted such that excursions in the blue wing have positive values. We see that the blue wing enhancements in \ion{Fe}{14} 264\AA{} are accompanied by a strong second emission component that has a fairly steady velocity at $\sim$50km/s from the first (core) component. This is consistent with what is observed for \ion{Fe}{14} 274\AA{}. We note also that the second component properties are remarkably consistent with those determined from the simple forward model presented in the previous section. In future work we hope to expand and explore this encouraging ``guided'' double component fit technique to more data sets of interest, but that is beyond the scope of the present effort. 

\section{Discussion}

Before we can close our analysis there are a couple of outstanding questions. First, could the oscillations in line width (of order 1.5 km/s) be caused by changes in the thermal broadening? In short, this is highly unlikely and for the following reason. In the locations we study in the current work, the overall line broadening ($\sigma_{1/e}$, the 1/e width) is of order 60 km/s, whereas the thermal width of these lines is of order 20 km/s. We know that the $\sigma_{1/e} = \sqrt{\sigma_{NT}^2 + \sigma_{inst}^2+ \sigma_{th}^2}$, with $\sigma_{inst}=22$m\AA{} the instrumental broadening, and $\sigma_{th}=\sqrt{2kT/m}$ with $k$ the Boltzmann constant and $m$ the ion mass. If we assume that the slight changes in line width $\Delta \sigma_{1/e}$ are caused only by thermal width changes, we can calculate the following:
\begin{equation}
  \sigma_{th}^2=\sigma_{1/e}^2-\sigma_{NT}^2-\sigma_{inst}^2 = 2 kT/m,
\end{equation}
and:
\begin{equation}
  \Delta T = m \sigma_{1/e} \Delta \sigma_{1/e} / k
\end{equation}

\begin{figure}
\epsscale{1.15}
\plotone{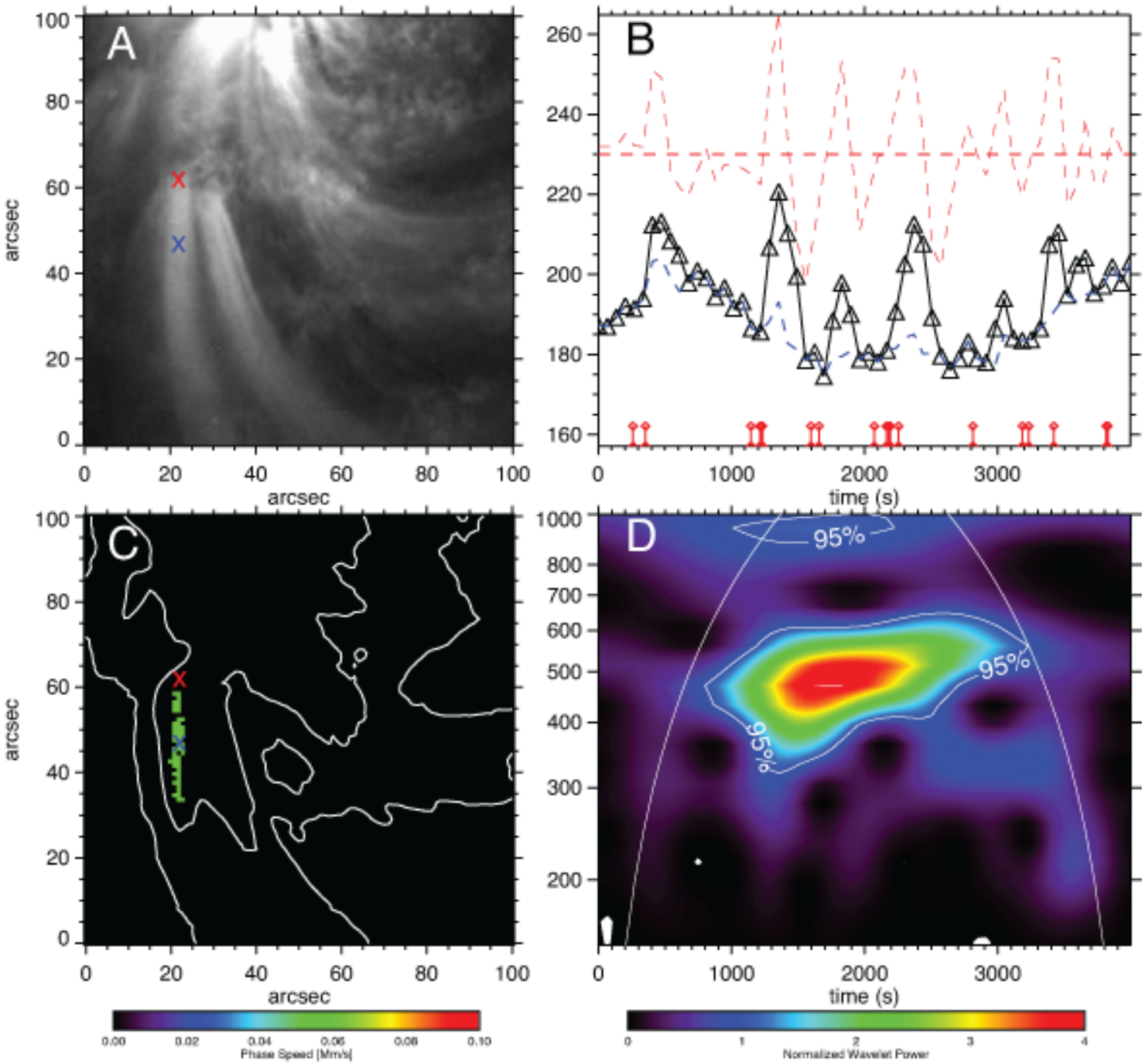}
\caption{Results of a Monte Carlo simulation illustrate that oscillation finding software cannot distinguish quasi-periodic signals arising from stochastically driven upflows from those of propagating waves. Panel (a) shows a {\em TRACE} 171\AA{} image where we added a series of simulated upflow events that originate at the footpoint of the cool loop emanating from the red cross. These upflow events are assigned the properties we expect for spicules heated to coronal temperatures: lifetimes of 100s, intensity of order 10\% of the background loop structure, apparent propagation speed of 70 km/s (chosen to be along the negative y-axis in this case), and slight widening and fading. Panel (b) shows the timing of the randomly chosen occurrence of the upflow events at the location of the red cross in panel (a) (red symbols along the bottom of the plot), with the original timeseries at the location of the blue cross in panel (a) (blue dashed line), the simulated timeseries (full black line), and its running time difference (red dashed). Note the slight time delay (because of the finite propagation speed) between the spicule occurrence at the red location, and the peaks in the timeseries at the blue location.  Panel (c) shows the phase speed (in Mm/s) for regions where the wavetracker found significant propagating signal in the 3.5 mHz (250 s) passband with white contours for intensity of the {\em TRACE} 171\AA{} images. Panel (d) shows the wavelet power for the timeseries of panel (b), i.e., for the blue cross location.  \label{f_tmc}}
\end{figure}

With the observed values of $\Delta \sigma_{1/e} \approx 1.5$ km/s, we estimate that $\Delta T \approx 0.66$ MK. Such a large change in temperature would significantly change the ionization balance, and essentially shift the temperature so far away from the peak of the contribution function for each ion, that very drastic changes in intensity would be expected. Because these are not observed, it is unlikely that the line width variations are predominantly caused by temperature changes associated with sound waves. This is further bolstered by the fact that the observed quasi-periodic variations are in phase for ions formed across a wide range of temperatures \citep[see e.g.,][for simulations of how temperature changes from sound waves can lead to strong phase shifts between oscillations of lines from ions in neighboring ionization stages]{2007SoPh..246..165P}. 

It seems from this analysis that  the line widths changes cannot be caused by thermal changes associated with sound waves that propagate on the structure responsible for the dominant emission. Is it possible that sound waves propagating on a steadily blueshifted component (compared to the dominant, stationary, one) can explain the observations? Such a scenario could in principle cause linewidth changes associated with blueward centroid shifts and intensity peaks. However, it would require a secondary component that is persistently blueshifted compared to the dominant component for the whole 3 hours of the timeseries. We believe that our analysis using both the R-B asymmetry variations and the double fits strongly implies the presence of a second component of emission that occurs at high velocity (i.e., it is associated with high velocity upflows). Our results thus pose a challenge to wave models to reproduce the observed amplitude of changes in line width, the clear R-B asymmetry variations, and the correlation of line widths with R-B signals at high velocities ($>$70km/s). Such models require sophisticated synthetic observables and numerical models of the solar atmosphere \citep[e.g.,][]{2007SoPh..246..165P} and are beyond the scope of the current paper.

The presence of a high-speed upflowing component is directly shown by the results of our study of double Gaussian fits to the time series in the previous section. Generally double fits for line profiles with such a weak secondary component are poorly constrained and the results of the fit are strongly affected by the initial guesses for the properties (intensity, velocity, width) of the second component \citep[][]{1998A&AS..132..145M}. The result is often highly noisy and so additional, strong physical or observational constraints must be applied - the approach attempted above directly uses observations, and is a first step in that direction. It is pertinent here to note the perils of single Gaussian fits to data such as those discussed above - those where a second emitting component is visible, either in the spectral or imaging data. Our analysis shows that the results of a single Gaussian fit to such data can significantly impact the physical interpretation in a quite unintentional way. It is possible that double fit analysis, driven from a knowledge of multiple emitting components in a resolution element, like that demonstrated here, can help reduce that ambiguity and potential misinterpretation. 

The results discussed in the sections above provide a compelling scenario in which upflows at high speed (of order 50-150 km/s)\footnote{The viewing angle will determine the amplitude of the line-of-sight velocity, determined from our RB, double fit and Monte Carlo analysis, and of the velocity in the plane of the sky, determined from space-time plots. The viewing angle likely changes from the footpoints to the upper parts of the loop.} and low brightness (of order a few percent of the background brightness) are responsible for quasi-periodic changes in line intensity, velocity, width, and blueward asymmetries that propagate away from magnetic footpoints along coronal loops. We do not claim that that this scenario can explain {\it all} reported observations of sound waves in the corona. After all, we are only observing one active region, and there are too many uncertainties (see below) at this point in time. However, our results strongly suggest that before the wave interpretation is applied to intensity oscillations, a careful analysis of higher order moments of spectra is required.

The presence of a second component that flows upward at speeds that are of order 50 km/s higher than the dominant plasma component begs the question how these two components are connected. Using the current data it is impossible to say whether the two components occur close to each other on neighboring field lines that are indistinguishable because of the spatio-temporal resolution of EIS which is modest compared to the fine-scale structuring and fast dynamics of the low solar atmosphere, or whether the enormous superposition of coronal lines along the line-of-sight samples two widely separated populations of plasma. 

In the following we briefly speculate on whether the ``flows'' scenario could explain some of the observed properties perceived to belong to slow magnetoacoustic waves \citep[for a review, see][]{2008JPhCS.118a2038N}? As one example, we suggest that the flows scenario can naturally explain the lack of clean oscillations in plage-related loops, and the predominance of non-periodic propagating disturbances that occur all over quiet Sun and plage-related loops \citep[][]{2009ApJ...707..524M,2009ApJ...706L..80M}. We note that these non-periodic propagating disturbances are identical in appearance (intensity, apparent propagation, lifetime) to those that are quasi-periodic and abound in the literature \citep[e.g.,][]{1999SoPh..187..261S, 2002ApJ...567L..89W,2007Sci...318.1585S, 2010arXiv1001.2022M, 2009ApJ...706L..80M, 2009ApJ...707..524M}. If the flows were driven stochastically by an as yet unknown mechanism that affects the magnetic field and occurs on dominant photospheric timescales (from $p$-mode oscillations at 4-10 minutes, and granulation on timescales of 5-15 minutes), one would not expect to see oscillations everywhere. In addition, the oscillations that {\it are} observed would not be expected to be very clean or significant. That is exactly what we see with {\em TRACE} and EIS. In addition, the timescale of quasi-periodicity would often vary strongly from location to location \citep[observed with {\em TRACE}, see][]{2002SoPh..209...61D,2002SoPh..209...89D}. In fact, detailed analysis of spectra in quiet Sun network regions shows that the blueward asymmetries do indeed sometimes recur quasi-periodically, but often do not, which is fully compatible with the {\em TRACE} observations \citep{2009ApJ...706L..80M}. We find similar behavior in the {\em EIS} data analyzed here. Fig.~\ref{f5} shows the results of a wavelet analysis that shows quasi-periodicities of order 5-12 minutes for all derived properties of the line profiles throughout the timeseries that can be readily compared to Fig.~14 of \citet{2009ApJ...707..524M}.

What drives these upflows? Give the proposed association with upflow events in the lower atmosphere \citep[e.g., spicules;][]{2009ApJ...706L..80M}, it is natural to assume a mechanism in which reconnection caused by photospheric motions which drive/alter the small-scale magnetic field in a stochastic manner. We should stress that the energy and mass release may not always lie in the chromosphere and appear as a spicule, but could well arise in the transition region. Given the range of observed photospheric motions and the prevalent timescales, quasi-periodicity would occur naturally. This quasi-periodicity can be accentuated because the observed coronal ``oscillations'' are typically identified using a wavelet analysis \citep[e.g.,][]{2007SoPh..241..397N,2008SoPh..248..395S}\footnote{In the specific case of the {\em EIS} data, it is possible that the pointing jitter perpendicular to the slit (which cannot be corrected for and can be of order 1-2\arcsec) sometimes acts as a secondary cause of some of the quasi-periodicities in Figs.\pref{f4}, \pref{f4b} and \pref{f4c}. We do not believe it is the dominant cause given the fact that such effects would impact the whole slit similarly throughout the whole timeseries. This is not what we observe \citep[see also,][]{2009A&A...503L..25W}}. 

To illustrate the ill-posedness of interpreting quasi-periodicities in the corona {\em using imaging data alone}, we have performed further forward modeling using Monte Carlo simulations. We start with a {\em TRACE} 171\AA{} image sequence and add an artificial signal that consists of a series of propagating events with properties derived from the analysis presented above: low intensity (10\% of background intensity), short lifetime at the footpoints (100s), high apparent propagation velocity in the plane of the sky (70km/s), and gradual fading and slight expansion in width (over e-folding distance of, respectively 15 and 20 Mm), as a result of decreasing density in the upflow events with height, e.g., because of fast expansion while the upflows merge into the pre-existing corona. 

To study the impact of a stochastic driver, we allow the ``heating'' events to occur randomly in time at the loop footpoint with a random uniform distribution given by the IDL function {$\tt randomu.pro$}. On average, we introduce $n$ events during the timeseries of duration $\Delta T$, with $\Delta T/n$ of order the granular timescale (e.g., 300s). We then use two analysis techniques that are commonly used to detect ``oscillations'' in coronal imaging data: a wavelet analysis of the timeseries of running differences (in this case with 120s time difference) in individual pixels \citep{1998BAMS...79...61T}, and our ``wavetracker'' software which focuses on detecting propagating signals in a set of narrow frequency ranges \citep{2008SoPh..252..321M}. An example of the analysis is given in Fig.~\pref{f_tmc} which shows the original {\em TRACE} 171 \AA{} image, the simulated and original timeseries in one location, and the results of both the wavetracker and wavelet analysis. We can see that in this particular case the random bunching in time of individual propagating heating events leads to a clearly detectable quasi-periodicity in the wavelet analysis, and a nicely propagating signal (with 70 km/s) in the wavetracker analysis. However, no waves are present in the original data, and no waves have been introduced in the simulated data. Only heating events with the properties that we have deduced from the EIS data are included. This underscores the severe limitations of oscillation finding techniques that are based on only intensity information, such as coronal images. 

To investigate how severe this issue is, we performed a large number of similar simulations with a range of characteristic timescales $\Delta T/n$ between 200 and 700s. This preliminary study shows that the combination of wavetracker and wavelet analysis of the resulting timeseries detects quasi-periodic signals (``oscillations'') in about 15-50\% of the realizations, depending on the amplitude of the heating events, the significance levels used in the wavelet analysis, and whether the running difference or original data is used for the wavelet analysis. This implies that the observed quasi-periodicities in the upflows are not necessarily a sign of a wave driver. More importantly, it illustrates that we cannot use imaging data alone to distinguish between upflow and wave scenarios, casting signficant doubt on the coronal seismology applications of ``slow-mode magneto-acoustic'' waves. Our results suggest that such studies will require careful analysis of at least the first three moments of simultaneously observed spectral line profiles.

Can the ``flows'' scenario explain what causes the rapid fading and uniquely upward apparent motions of the propagating disturbances? In the wave scenario, this has previously been attributed to damping of waves \citep{2006SoPh..234...41K}. In our scenario the visibility of the upward flows would critically depend on the density of the disturbances and on how the preexisting coronal plasma absorbs these mass injections. Without a proper model of what drives these upflows, the associated heating and the impact on the corona, it is very difficult to predict how the density (and velocity!) of these upflow events would change with height. Perhaps the coronal part of the upflow events tracks the behavior of the chromospheric counterpart (type II spicules) which show a clear decrease of density with height in {\em Hinode}/SOT \ion{Ca}{2} H timeseries \citep{2007PASJ...59S.655D}? In other words, perhaps the rapid decrease of density with height in the upflow events is responsible for the apparent fading of the disturbances with distance along the coronal loops? Such a fading could perhaps also explain the fact that \citet{2009A&A...503L..25W} found longer ``periods'' higher in the loop fan than at the footpoint. If only a fraction of the upflow events were dense enough to be visible at distances of 20~Mm from the loop footpoints, the recurrence of the motions observed higher in the corona would then tend to be on longer timescales, since some of the events recurring on shorter timescales would be invisible far away from the loop footpoints. 

An interesting twist to this scenario is added by preliminary analysis of AIA and Hinode data of upflow events, which indicates that the upflows are associated with heating of some rapidly moving chromospheric plasma to coronal temperatures. It is clear that such rapid heating could, in principle, lead to the rapid propagation into the corona of both thermal conduction fronts and/or sound waves. Depending on the initial conditions and details of the heating events, thermal conduction fronts can propagate even faster than the speed of sound \citep[e.g.,][]{Hansteen1993}. Perhaps the strong heating and strong acceleration associated with these heating events then leads to the presence of both flows and associated thermal conduction front/sound waves? We believe that in the dataset considered here the flows dominate the observed oscillations, but it is clear that careful analysis of spectra, and advanced numerical modeling will be required to disentangle which effects dominate in each observation.



The impact of the results discussed in this Paper are wide-ranging. Coronal seismology based on the waves interpretation of these propagating disturbances in coronal loops has been used to investigate the isothermal nature of loops, thermal substructuring of loops, the leakage of waves from the photosphere, the nature of thermal conductivity, etc \citep[e.g.,][]{2001A&A...370..591R, 2002SoPh..209...89D, 2002SoPh..209...61D, 2003A&A...404L...1K, 2007SoPh..246...53D, 2008SoPh..252..101D, 2008ApJ...681L..41M, 2008A&A...487L..17V, 2008A&A...482L...9O, 2009A&A...494..339O, 2009SSRv..tmp...38D, 2009ApJ...696.1448W, 2009ApJ...699L..72D, 2009ApJ...697.1674M, 2009ApJ...706L..76M}. Indeed, a similar slow-mode MHD wave interpretation exists for high velocity propagating disturbances of small amplitude observed on ubiquitous features observed in polar coronal holes, polar plumes \citep[see, e.g.,][to name only a few of the many papers on the topic]{1998ApJ...501L.217D, 1999ApJ...514..441O, 2000SoPh..196...63B, 2009A&A...499L..29B}. We believe that our results may affect the interpretation of polar plume disturbances as well. \citet{2010A&A...510L...2M} showed that the similarity in propagation speeds, amplitudes and quasi-periodicities of disturbances on polar plumes and those observed in other magnetic locations of the solar atmosphere suggests that the motions observed on plumes may not be the result of the passage of MHD waves, but a consequence of the quasi-periodic appearance of driven upflows. While we have not yet performed the kind of detailed spectroscopic analysis presented here on a polar plume (or an equivalent in an equatorial coronal hole, which would provide the best opportunity for detecting profile asymmetries), it seems that the (almost global) properties and ubiquity of these upflow events indicates that a similar mechanism may be at play at the roots of the fast solar wind. This is illustrated by the {\em SOHO} coronal hole observation analysis of \citet{2009ApJ...701L...1D} (and that of an upcoming paper \-- McIntosh, Leamon \& De Pontieu, ``The Spectroscopic Footprint of the Fast Solar Wind'', submitted to the ApJ).

While a more extensive statistical study of oscillations using spectroscopic data is required, it is clear that if our interpretation of the data presented is correct (i.e., that some of the disturbances are the result of driven upflows originating in the lower solar atmosphere), then much of the coronal seismology effort based on the interpretation of longitudinal, compressive, slow MHD modes in coronal regions not directly rooted in sunspots has to be reconsidered. The upflow interpretation on the other hand opens a new window into the connection between the chromosphere and corona with significant implications for coronal heating mechanisms \citep[e.g.,][]{2007PASJ...59S.655D}.

\section{Conclusion}
We have demonstrated that recent EIS measurements of intensity and velocity oscillations, interpreted as direct evidence for propagating slow-mode waves in the corona, are actually accompanied by oscillations in the line width, and recurring asymmetries in line profiles across a range of temperatures. These facts, at the least, imply that coronal seismology using such propagating disturbances in the corona is not as straightforward as assumed. Moreover, we show that these changes in intensity, velocity, line width and blueward asymmetry are compatible with a scenario in which faint upflows at high speed occur quasi-periodically and cause oscillations in the parameters determined from single Gaussian fits to spectral line profiles. Our results indicate that a significant fraction of the quasi-periodicities observed with coronal imagers and spectrographs that have previously been interpreted as propagating magnetoacoustic waves, may instead be caused by these quasi-periodic upflows. The uncertainty in the identification of the physical cause for coronal oscillations significantly impacts the prospects of successful coronal seismology using propagating, slow-mode magneto-acoustic waves. At the same time, the association of these propagating disturbances in coronal loops with high speed upflows provides an exciting new window into the mechanism that propels hot plasma into the corona.

\acknowledgements
This work is supported by NASA grants NNX08AL22G and NNX08BA99G to SWM and BDP. The National Center for Atmospheric Research is sponsored by the National Science Foundation. Thanks to Alan Title for discussions.



\begin{thebibliography}{54}
\expandafter\ifx\csname natexlab\endcsname\relax\def\natexlab#1{#1}\fi

\bibitem[{{Andries} {et~al.}(2005){Andries}, {Arregui}, \&
  {Goossens}}]{2005ApJ...624L..57A}
{Andries}, J., {Arregui}, I., \& {Goossens}, M. 2005, \apj, 624, L57

\bibitem[{{Banerjee} {et~al.}(2000){Banerjee}, {O'Shea}, \&
  {Doyle}}]{2000SoPh..196...63B}
{Banerjee}, D., {O'Shea}, E., \& {Doyle}, J.~G. 2000, \solphys, 196, 63

\bibitem[{{Banerjee} {et~al.}(2009){Banerjee}, {Teriaca}, {Gupta}, {Imada},
  {Stenborg}, \& {Solanki}}]{2009A&A...499L..29B}
{Banerjee}, D., {Teriaca}, L., {Gupta}, G.~R., {Imada}, S., {Stenborg}, G., \&
  {Solanki}, S.~K. 2009, \aap, 499, L29

\bibitem[{{Bevington} \& {Robinson}(2003)}]{2003drea.book.....B}
{Bevington}, P.~R., \& {Robinson}, D.~K. 2003, {Data reduction and error
  analysis for the physical sciences} (Data reduction and error analysis for
  the physical sciences, 3rd ed., by Philip R.~Bevington, and Keith
  D.~Robinson.~Boston, MA: McGraw-Hill, ISBN 0-07-247227-8, 2003.)

\bibitem[{{Brown} {et~al.}(2008){Brown}, {Feldman}, {Seely}, {Korendyke}, \&
  {Hara}}]{2008ApJS..176..511B}
{Brown}, C.~M., {Feldman}, U., {Seely}, J.~F., {Korendyke}, C.~M., \& {Hara},
  H. 2008, \apjs, 176, 511

\bibitem[{{Culhane} {et~al.}(2007){Culhane}, {Harra}, {James}, {Al-Janabi},
  {Bradley}, {Chaudry}, {Rees}, {Tandy}, {Thomas}, {Whillock}, {Winter},
  {Doschek}, {Korendyke}, {Brown}, {Myers}, {Mariska}, {Seely}, {Lang}, {Kent},
  {Shaughnessy}, {Young}, {Simnett}, {Castelli}, {Mahmoud}, {Mapson-Menard},
  {Probyn}, {Thomas}, {Davila}, {Dere}, {Windt}, {Shea}, {Hagood}, {Moye},
  {Hara}, {Watanabe}, {Matsuzaki}, {Kosugi}, {Hansteen}, \&
  {Wikstol}}]{2007SoPh..243...19C}
{Culhane}, J.~L., {et~al.} 2007, \solphys, 243, 19

\bibitem[{{De Moortel}(2009)}]{2009SSRv..tmp...38D}
{De Moortel}, I. 2009, Space Science Reviews, 38

\bibitem[{{De Moortel} \& {Bradshaw}(2008)}]{2008SoPh..252..101D}
{De Moortel}, I., \& {Bradshaw}, S.~J. 2008, \solphys, 252, 101

\bibitem[{{de Moortel} {et~al.}(2002{\natexlab{a}}){de Moortel}, {Hood},
  {Ireland}, \& {Walsh}}]{2002SoPh..209...89D}
{de Moortel}, I., {Hood}, A.~W., {Ireland}, J., \& {Walsh}, R.~W.
  2002{\natexlab{a}}, \solphys, 209, 89

\bibitem[{{de Moortel} {et~al.}(2000){de Moortel}, {Ireland}, \&
  {Walsh}}]{2000A&A...355L..23D}
{de Moortel}, I., {Ireland}, J., \& {Walsh}, R.~W. 2000, \aap, 355, L23

\bibitem[{{de Moortel} {et~al.}(2002{\natexlab{b}}){de Moortel}, {Ireland},
  {Walsh}, \& {Hood}}]{2002SoPh..209...61D}
{de Moortel}, I., {Ireland}, J., {Walsh}, R.~W., \& {Hood}, A.~W.
  2002{\natexlab{b}}, \solphys, 209, 61

\bibitem[{{De Moortel} \& {Pascoe}(2009)}]{2009ApJ...699L..72D}
{De Moortel}, I., \& {Pascoe}, D.~J. 2009, \apjl, 699, L72

\bibitem[{{de Moortel} \& {Rosner}(2007)}]{2007SoPh..246...53D}
{de Moortel}, I., \& {Rosner}, R. 2007, \solphys, 246, 53

\bibitem[{{De Pontieu} {et~al.}(2009){De Pontieu}, {McIntosh}, {Hansteen}, \&
  {Schrijver}}]{2009ApJ...701L...1D}
{De Pontieu}, B., {McIntosh}, S.~W., {Hansteen}, V.~H., \& {Schrijver}, C.~J.
  2009, \apjl, 701, L1

\bibitem[{{de Pontieu} {et~al.}(2007){de Pontieu}, {McIntosh}, {Hansteen},
  {Carlsson}, {Schrijver}, {Tarbell}, {Title}, {Shine}, {Suematsu}, {Tsuneta},
  {Katsukawa}, {Ichimoto}, {Shimizu}, \& {Nagata}}]{2007PASJ...59S.655D}
{de Pontieu}, B., {et~al.} 2007, \pasj, 59, 655

\bibitem[{{Deforest} \& {Gurman}(1998)}]{1998ApJ...501L.217D}
{Deforest}, C.~E., \& {Gurman}, J.~B. 1998, \apj, 501, L217+

\bibitem[{{Fleck} {et~al.}(1995){Fleck}, {Domingo}, \&
  {Poland}}]{1995somi.book.....F}
{Fleck}, B., {Domingo}, V., \& {Poland}, A. 1995, {The SOHO mission}, ed.
  {Fleck, B., Domingo, V., \& Poland, A.}

\bibitem[{{Handy} {et~al.}(1999){Handy}, {Acton}, {Kankelborg}, {Wolfson},
  {Akin}, {Bruner}, {Caravalho}, {Catura}, {Chevalier}, {Duncan}, {Edwards},
  {Feinstein}, {Freeland}, {Friedlaender}, {Hoffmann}, {Hurlburt}, {Jurcevich},
  {Katz}, {Kelly}, {Lemen}, {Levay}, {Lindgren}, {Mathur}, {Meyer}, {Morrison},
  {Morrison}, {Nightingale}, {Pope}, {Rehse}, {Schrijver}, {Shine}, {Shing},
  {Strong}, {Tarbell}, {Title}, {Torgerson}, {Golub}, {Bookbinder}, {Caldwell},
  {Cheimets}, {Davis}, {Deluca}, {McMullen}, {Warren}, {Amato}, {Fisher},
  {Maldonado}, \& {Parkinson}}]{1999SoPh..187..229H}
{Handy}, B.~N., {et~al.} 1999, \solphys, 187, 229

\bibitem[Hansteen(1993)]{Hansteen1993} Hansteen, V.\ 1993, \apj, 402, 741 

\bibitem[{{Hara} {et~al.}(2008){Hara}, {Watanabe}, {Harra}, {Culhane}, {Young},
  {Mariska}, \& {Doschek}}]{2008ApJ...678L..67H}
{Hara}, H., {Watanabe}, T., {Harra}, L.~K., {Culhane}, J.~L., {Young}, P.~R.,
  {Mariska}, J.~T., \& {Doschek}, G.~A. 2008, \apjl, 678, L67

\bibitem[{{Howard} {et~al.}(2008){Howard}, {Moses}, {Vourlidas}, {Newmark},
  {Socker}, {Plunkett}, {Korendyke}, {Cook}, {Hurley}, {Davila}, {Thompson},
  {St Cyr}, {Mentzell}, {Mehalick}, {Lemen}, {Wuelser}, {Duncan}, {Tarbell},
  {Wolfson}, {Moore}, {Harrison}, {Waltham}, {Lang}, {Davis}, {Eyles},
  {Mapson-Menard}, {Simnett}, {Halain}, {Defise}, {Mazy}, {Rochus}, {Mercier},
  {Ravet}, {Delmotte}, {Auchere}, {Delaboudiniere}, {Bothmer}, {Deutsch},
  {Wang}, {Rich}, {Cooper}, {Stephens}, {Maahs}, {Baugh}, {McMullin}, \&
  {Carter}}]{2008SSRv..136...67H}
{Howard}, R.~A., {et~al.} 2008, Space Science Reviews, 136, 67

\bibitem[{{King} {et~al.}(2003){King}, {Nakariakov}, {Deluca}, {Golub}, \&
  {McClements}}]{2003A&A...404L...1K}
{King}, D.~B., {Nakariakov}, V.~M., {Deluca}, E.~E., {Golub}, L., \&
  {McClements}, K.~G. 2003, \aap, 404, L1

\bibitem[{{Klimchuk}(2006)}]{2006SoPh..234...41K}
{Klimchuk}, J.~A. 2006, \solphys, 234, 41

\bibitem[{{Kosugi} {et~al.}(2007){Kosugi}, {Matsuzaki}, {Sakao}, {Shimizu},
  {Sone}, {Tachikawa}, {Hashimoto}, {Minesugi}, {Ohnishi}, {Yamada}, {Tsuneta},
  {Hara}, {Ichimoto}, {Suematsu}, {Shimojo}, {Watanabe}, {Shimada}, {Davis},
  {Hill}, {Owens}, {Title}, {Culhane}, {Harra}, {Doschek}, \&
  {Golub}}]{2007SoPh..243....3K}
{Kosugi}, T., {et~al.} 2007, \solphys, 243, 3

\bibitem[{{Mariska} \& {Muglach}(2010)}]{2010arXiv1003.0420M}
{Mariska}, J.~T., \& {Muglach}, K. 2010, ArXiv e-prints

\bibitem[{{Mariska} {et~al.}(2008){Mariska}, {Warren}, {Williams}, \&
  {Watanabe}}]{2008ApJ...681L..41M}
{Mariska}, J.~T., {Warren}, H.~P., {Williams}, D.~R., \& {Watanabe}, T. 2008,
  \apjl, 681, L41

\bibitem[{{Marsh} \& {Walsh}(2009)}]{2009ApJ...706L..76M}
{Marsh}, M.~S., \& {Walsh}, R.~W. 2009, \apjl, 706, L76

\bibitem[{{Marsh} {et~al.}(2009){Marsh}, {Walsh}, \&
  {Plunkett}}]{2009ApJ...697.1674M}
{Marsh}, M.~S., {Walsh}, R.~W., \& {Plunkett}, S. 2009, \apj, 697, 1674

\bibitem[{{Mazzotta} {et~al.}(1998){Mazzotta}, {Mazzitelli}, {Colafrancesco},
  \& {Vittorio}}]{1998A&AS..133..403M}
{Mazzotta}, P., {Mazzitelli}, G., {Colafrancesco}, S., \& {Vittorio}, N. 1998,
  \aaps, 133, 403

\bibitem[{{McIntosh} \& {De Pontieu}(2009{\natexlab{a}})}]{2009ApJ...707..524M}
{McIntosh}, S.~W., \& {De Pontieu}, B. 2009{\natexlab{a}}, \apj, 707, 524

\bibitem[{{McIntosh} \& {De Pontieu}(2009{\natexlab{b}})}]{2009ApJ...706L..80M}
---. 2009{\natexlab{b}}, \apjl, 706, L80

\bibitem[{{McIntosh} {et~al.}(2010{\natexlab{a}}){McIntosh}, {De Pontieu}, \&
  {Leamon}}]{2010arXiv1001.2022M}
{McIntosh}, S.~W., {De Pontieu}, B., \& {Leamon}, R.~J. 2010{\natexlab{a}},
  ArXiv e-prints

\bibitem[{{McIntosh} {et~al.}(2008){McIntosh}, {de Pontieu}, \&
  {Tomczyk}}]{2008SoPh..252..321M}
{McIntosh}, S.~W., {de Pontieu}, B., \& {Tomczyk}, S. 2008, \solphys, 252, 321

\bibitem[{{McIntosh} {et~al.}(1998){McIntosh}, {Diver}, {Judge}, {Charbonneau},
  {Ireland}, \& {Brown}}]{1998A&AS..132..145M}
{McIntosh}, S.~W., {Diver}, D.~A., {Judge}, P.~G., {Charbonneau}, P.,
  {Ireland}, J., \& {Brown}, J.~C. 1998, \aaps, 132, 145

\bibitem[{{McIntosh} {et~al.}(2010{\natexlab{b}}){McIntosh}, {Innes}, {de
  Pontieu}, \& {Leamon}}]{2010A&A...510L...2M}
{McIntosh}, S.~W., {Innes}, D.~E., {de Pontieu}, B., \& {Leamon}, R.~J.
  2010{\natexlab{b}}, \aap, 510, L2+

\bibitem[{{Nakariakov}(2008)}]{2008JPhCS.118a2038N}
{Nakariakov}, V.~M. 2008, Journal of Physics Conference Series, 118, 012038

\bibitem[{{Nakariakov} \& {King}(2007)}]{2007SoPh..241..397N}
{Nakariakov}, V.~M., \& {King}, D.~B. 2007, \solphys, 241, 397

\bibitem[{{Nakariakov} \& {Verwichte}(2005)}]{2005LRSP....2....3N}
{Nakariakov}, V.~M., \& {Verwichte}, E. 2005, Living Reviews in Solar Physics,
  2, 3

\bibitem[{{Ofman} {et~al.}(1999){Ofman}, {Nakariakov}, \&
  {Deforest}}]{1999ApJ...514..441O}
{Ofman}, L., {Nakariakov}, V.~M., \& {Deforest}, C.~E. 1999, \apj, 514, 441

\bibitem[{{Ofman} \& {Wang}(2008)}]{2008A&A...482L...9O}
{Ofman}, L., \& {Wang}, T.~J. 2008, \aap, 482, L9

\bibitem[{{Owen} {et~al.}(2009){Owen}, {De Moortel}, \&
  {Hood}}]{2009A&A...494..339O}
{Owen}, N.~R., {De Moortel}, I., \& {Hood}, A.~W. 2009, \aap, 494, 339

\bibitem[{{Pascoe} {et~al.}(2007){Pascoe}, {Nakariakov}, \&
  {Arber}}]{2007SoPh..246..165P}
{Pascoe}, D.~J., {Nakariakov}, V.~M., \& {Arber}, T.~D. 2007, \solphys, 246,
  165

\bibitem[{{Peter}(2000)}]{2000A&A...360..761P}
{Peter}, H. 2000, \aap, 360, 761

\bibitem[{{Robbrecht} {et~al.}(2001){Robbrecht}, {Verwichte}, {Berghmans},
  {Hochedez}, {Poedts}, \& {Nakariakov}}]{2001A&A...370..591R}
{Robbrecht}, E., {Verwichte}, E., {Berghmans}, D., {Hochedez}, J.~F., {Poedts},
  S., \& {Nakariakov}, V.~M. 2001, \aap, 370, 591

\bibitem[{{Rouppe van der Voort} {et~al.}(2009){Rouppe van der Voort},
  {Leenaarts}, {de Pontieu}, {Carlsson}, \& {Vissers}}]{2009ApJ...705..272R}
{Rouppe van der Voort}, L., {Leenaarts}, J., {de Pontieu}, B., {Carlsson}, M.,
  \& {Vissers}, G. 2009, \apj, 705, 272

\bibitem[{{Sakao} {et~al.}(2007){Sakao}, {Kano}, {Narukage}, {Kotoku}, {Bando},
  {DeLuca}, {Lundquist}, {Tsuneta}, {Harra}, {Katsukawa}, {Kubo}, {Hara},
  {Matsuzaki}, {Shimojo}, {Bookbinder}, {Golub}, {Korreck}, {Su}, {Shibasaki},
  {Shimizu}, \& {Nakatani}}]{2007Sci...318.1585S}
{Sakao}, T., {et~al.} 2007, Science, 318, 1585

\bibitem[{{Schrijver} {et~al.}(1999){Schrijver}, {Title}, {Berger}, {Fletcher},
  {Hurlburt}, {Nightingale}, {Shine}, {Tarbell}, {Wolfson}, {Golub},
  {Bookbinder}, {Deluca}, {McMullen}, {Warren}, {Kankelborg}, {Handy}, \& {de
  Pontieu}}]{1999SoPh..187..261S}
{Schrijver}, C.~J., {et~al.} 1999, \solphys, 187, 261

\bibitem[{{Sych} \& {Nakariakov}(2008)}]{2008SoPh..248..395S}
{Sych}, R.~A., \& {Nakariakov}, V.~M. 2008, \solphys, 248, 395

\bibitem[{{Torrence} \& {Compo}(1998)}]{1998BAMS...79...61T}
{Torrence}, C., \& {Compo}, G.~P. 1998, Bulletin of the American Meteorological
  Society, 79, 61

\bibitem[{{Van Doorsselaere} {et~al.}(2008{\natexlab{a}}){Van Doorsselaere},
  {Brady}, {Verwichte}, \& {Nakariakov}}]{2008A&A...491L...9V}
{Van Doorsselaere}, T., {Brady}, C.~S., {Verwichte}, E., \& {Nakariakov}, V.~M.
  2008{\natexlab{a}}, \aap, 491, L9

\bibitem[{{Van Doorsselaere} {et~al.}(2008{\natexlab{b}}){Van Doorsselaere},
  {Nakariakov}, {Young}, \& {Verwichte}}]{2008A&A...487L..17V}
{Van Doorsselaere}, T., {Nakariakov}, V.~M., {Young}, P.~R., \& {Verwichte}, E.
  2008{\natexlab{b}}, \aap, 487, L17

\bibitem[{{Wang} {et~al.}(2009{\natexlab{a}}){Wang}, {Ofman}, \&
  {Davila}}]{2009ApJ...696.1448W}
{Wang}, T.~J., {Ofman}, L., \& {Davila}, J.~M. 2009{\natexlab{a}}, \apj, 696,
  1448

\bibitem[{{Wang} {et~al.}(2009{\natexlab{b}}){Wang}, {Ofman}, {Davila}, \&
  {Mariska}}]{2009A&A...503L..25W}
{Wang}, T.~J., {Ofman}, L., {Davila}, J.~M., \& {Mariska}, J.~T.
  2009{\natexlab{b}}, \aap, 503, L25

\bibitem[{{Wang} {et~al.}(2008){Wang}, {Solanki}, \&
  {Selwa}}]{2008A&A...489.1307W}
{Wang}, T.~J., {Solanki}, S.~K., \& {Selwa}, M. 2008, \aap, 489, 1307

\bibitem[{{Winebarger} {et~al.}(2002){Winebarger}, {Warren}, {van
  Ballegooijen}, {DeLuca}, \& {Golub}}]{2002ApJ...567L..89W}
{Winebarger}, A.~R., {Warren}, H., {van Ballegooijen}, A., {DeLuca}, E.~E., \&
  {Golub}, L. 2002, \apjl, 567, L89

\end{thebibliography}


\newpage

\appendix
In the appendix we show two more examples (Figs.~\pref{f4b} and \pref{f4c}) of correlated oscillations in intensity, line centroid, linewidth, and blueward asymmetries. Here we did not sum over 5 pixels (as in Fig.~\pref{f4}), but over just two pixels. The locations where we have found the oscillations are in the same region as the two examples shown by \citet{2009A&A...503L..25W}.

\begin{figure}
\epsscale{.75}
\plotone{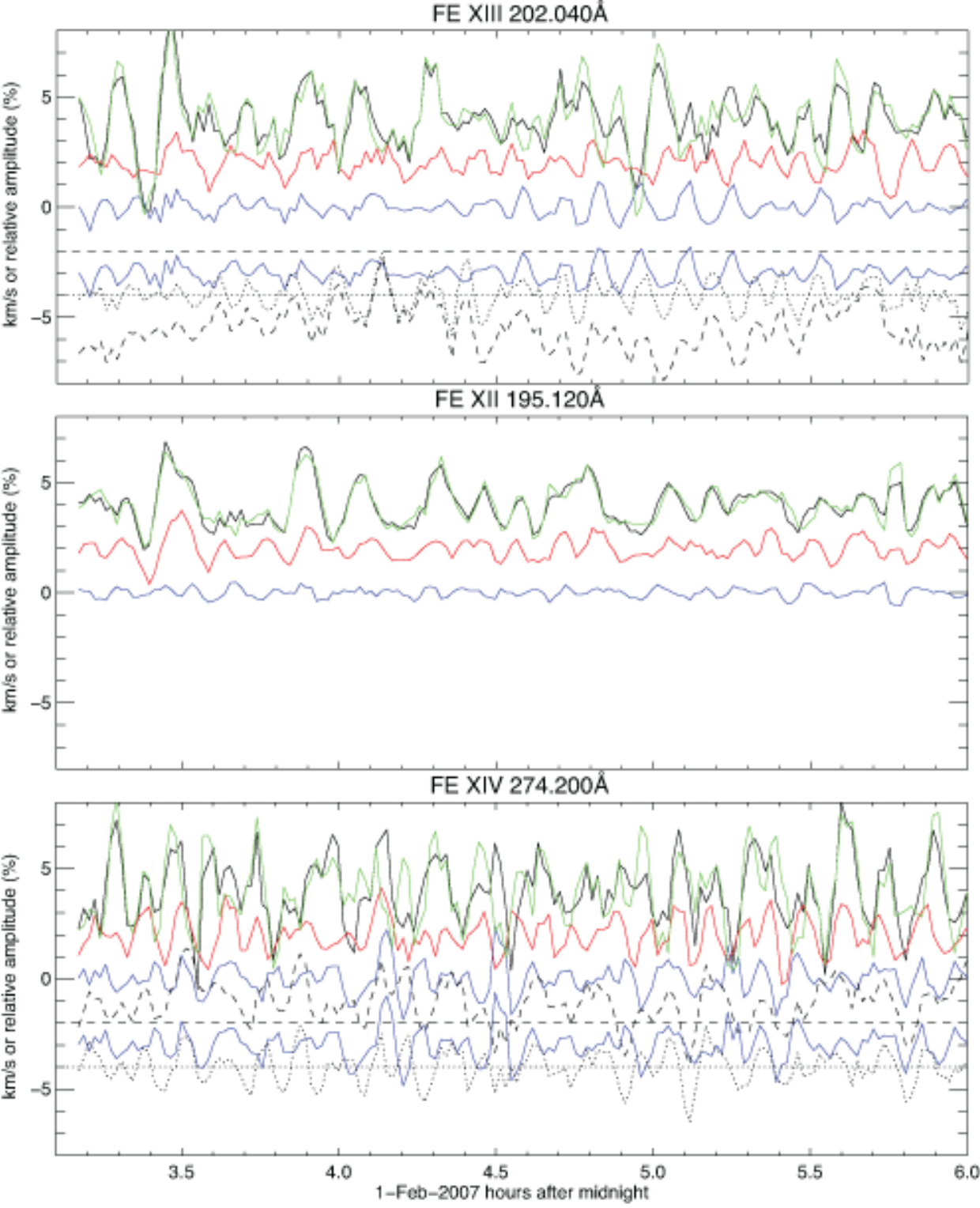}
\caption{Extracted EIS timeseries for the \ion{Fe}{13} 202\AA{} (top), \ion{Fe}{12}195\AA{} (middle) and \ion{Fe}{14} 274\AA{} (bottom) emission lines from 03:06-06:00UT. This figure is similar to Fig.~\ref{f4}, but for a different location (y=148\arcsec). In each panel we show the detrended percentage change in signal amplitude or velocity change in km/s. The green trace shows the temporal variance in the peak line intensity while the black, red, and blue traces show the changes in total profile intensity, Doppler velocity (line center shift), and line width respectively. Compare and contrast with the black dashed and dotted lines which show the 74-157km/s R-B analysis and detrended form (as for V and V$_{nt}$). The zero lines for these are drawn at -2 and -4 respectively and to stress our convention that {\em values above that zero line represent an excess in the blue wing of the line}. For clarity, the timeseries for intensity is offset by +4, velocity by +2 and the bottom linewidth by -2 from the zero-point on the y-axis. The presence of a second component in this location is also indicated in Fig. \ref{f7}. \label{f4b}}
\end{figure}
\begin{figure}
\plotone{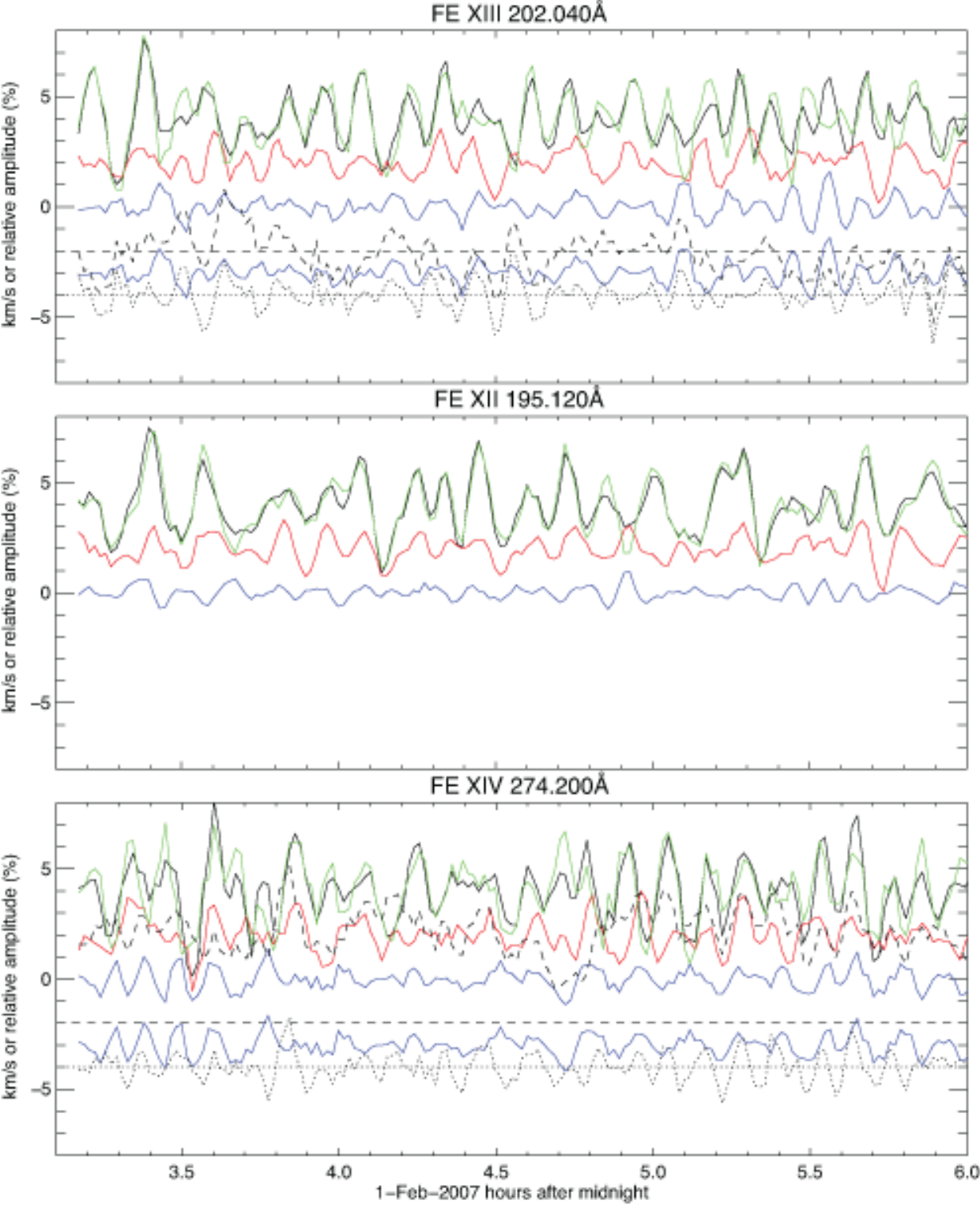}
\caption{Extracted EIS timeseries for the \ion{Fe}{13} 202\AA{} (top), \ion{Fe}{12}195\AA{} (middle) and \ion{Fe}{14} 274\AA{} (bottom) emission lines from 03:06-06:00UT. This figure is similar to Fig.~\ref{f4}, but for a different location (y=106\arcsec). In each panel we show the detrended percentage change in signal amplitude or velocity change in km/s. The green trace shows the temporal variance in the peak line intensity while the black, red, and blue traces show the changes in total profile intensity, Doppler velocity (line center shift), and line width respectively. Compare and contrast with the black dashed and dotted lines which show the 74-157km/s R-B analysis and detrended form (as for V and V$_{nt}$). The zero lines for these are drawn at -2 and -4 respectively and to stress our convention that {\em values above that zero line represent an excess in the blue wing of the line}. For clarity, the timeseries for intensity is offset by +4, velocity by +2 and the bottom linewidth by -2 from the zero-point on the y-axis. \label{f4c}}
\end{figure}

\end{document}